\documentclass[11pt,aps,prd,onecolumn,eqsecnum,nofootinbib,showkeys]       
{revtex4-2}
\usepackage{graphicx,epsfig,color}
\usepackage{bm}

\newcommand{\bc}{\begin{center}}
\newcommand{\ec}{\end{center}}
\newcommand{\be}{\begin{equation}}
\newcommand{\ee}{\end{equation}}
\newcommand{\bea}{\begin{eqnarray}}
\newcommand{\eea}{\end{eqnarray}}
\newcommand{\ba}{\begin{array}}
\newcommand{\ea}{\end{array}}
\newcommand{\lb}{\label}
\newcommand{\rf}{\ref}
\newcommand{\bfg}{\begin{figure}[htbp]}
\newcommand{\efg}{\end{figure}}

\begin{document}

\title{Meson scattering and tetraquarks \protect \\
in two-dimensional QCD}


\author{Hagop Sazdjian}
\email[]{hagop.sazdjian@ijclab.in2p3.fr}
\affiliation{Universit\'e Paris-Saclay, CNRS/IN2P3, IJCLab, 91405
Orsay, France}



\begin{abstract}
Two-quark--two-antiquark systems with four different quark flavors
are considered in the framework of two-dimensional QCD, in the
light-cone gauge, at leading orders of the $1/N_c^{}$ expansion.
Introducing basis functions with color-singlet mesonic clusters,
integral equations are established for the Green's functions and
the related scattering amplitudes involved in the sectors of the
direct and quark-exchange channels.
The problem of infrared divergences is dealt with via a systematic use
of an infrared regulator cutoff introduced in the gluon propagator.
It is shown that the on-mass shell scattering amplitudes are free of
infrared divergences up to order $1/N_c^2$. In the limit of
vanishing of the infrared cutoff, they can be represented by
effective four-meson contact-type interaction terms with unitarity
loops, calculable in terms of the meson wave functions and
propagators.
The results, obtained at order $1/N_c^2$, are summed with the
constraint of unitarization. The unitarized scattering amplitudes
are continued in the total mass-squared variable below the
two-meson thresholds, leading to a tetraquark bound state equation,
which generally has one solution. Spectroscopic applications of the
above results are sketched and discussed.   
\end{abstract}


\maketitle

\section{Introduction} \lb{s1}

The experimetal discovery, during the last two decades, of many
candidates of ``exotic hadrons''
\cite{Choi:2003ue,Aubert:2003fg,Besson:2003cp,Aubert:2005rm,
Ablikim:2013mio,Liu:2013dau,Ablikim:2013wzq,Aaij:2014jqa,Aaij:2015tga,
Aaij:2020fnh,Aaij:2020ypa,Wu:2020hmk,LHCb:2021auc},
which are assumed to contain
more valence quarks than the ordinary mesons and baryons
\cite{GellMann:1964nj,Zweig:1964jf}, has led to intense investigations
about the analysis of their internal structures 
\cite{Maiani:2004vq,Maiani:2005pe,Ebert:2007rn,Chen:2016qju,
Hosaka:2016pey,Lebed:2016hpi,Esposito:2016noz,Ali:2017jda,
Guo:2017jvc,Olsen:2017bmm,Karliner:2017qjm,Karliner:2017qhf,
Eichten:2017ffp,Albuquerque:2018jkn,Albuquerque:2021tqd,Liu:2019zoy,
Ali:2019roi,Brambilla:2019esw,Prelovsek:2020emb,Alexandrou:2023cqg,
Radhakrishnan:2024ihu,Lucha:2021mwx,
Heupel:2012ua,Eichmann:2020oqt,Hoffer:2024fgm,Berwein:2024ztx}.
Contrary to ordinary hadrons, exotic hadrons have the
possibility of being internally decomposed into ordinary hadronic
clusters, which could then screen the action of the confining forces
between quarks and could deform the expected compact
structure into a loosely bound molecular-type structure
\cite{Jaffe:2008zz,Weinstein:1982gc,Wang:1992wi,Nielsen:2009uh,
Lucha:2019cdc,Sazdjian:2022kaf}.
\par
The latter phenomenon complicates the task of analyzing
the properties of the exotic hadrons. The main reason of this  comes
from our incomplete control of the confining forces. QCD is a
nonperturbative theory in the infrared region, or, equivalently, at
large distances, where it becomes confining. The tools that we
dispose of are either empirical, or approximate, or at best numerical,
with severe limitations. Conventional methods that hinge on the use
of additive confining potentials, or additive empirical gluon
propagators for the study of the many-body bound state problem,
have the drawback of producing residual long-range van der Waals
forces that are not present on experimental grounds
\cite{Fishbane:1977ay,Appelquist:1978rt,Willey:1978fm,Matsuyama:1978hf,
Gavela:1979zu,Lenz:1985jk} and that, therefore, might alter some of
the qualitative features of the problem under study.
\par
From this point of view, two-dimensional QCD in the large-$N_c^{}$
limit, where $N_c^{}$ is the parameter of the color-gauge group
SU($N_c^{}$), which was first introduced and studied by 't~Hooft
\cite{'tHooft:1973jz,'tHooft:1974hx}, has been revealed as an
efficient tool for probing many of the problems related with the
confinement of quarks. In two dimensions, confinement is a basic
property of the theory with controllable infrared behavior. The
large-$N_c^{}$ limit ensures the damping of all inelasticity and
pair creation effects and reduces the class of dominant Feynman
diagrams to that of ``planar'' ones for color-singlet irreducible
systems
\cite{'tHooft:1973jz,Witten:1979kh,Witten:1979pi,Coleman:1985rnk,
tHooft:2002ufq}.
In noncovariant gauges \cite{Leibbrandt:1987qv} in two dimensions,
trilinear and quadrilinear gluon couplings disappear and the
single gluon propagator becomes the only dynamical object of the
theory; in particular, in the many-body case, the interactions
become free of the residual long-range van der Waals forces.
Two-dimensional QCD at large $N_c^{}$ has
later been considered by many authors in its various aspects related
with confinement \cite{Callan:1975ps,Einhorn:1976uz,Einhorn:1976ax,
Einhorn:1977bg,Hanson:1976ey,Hildebrandt:1977ny,Bars:1977ud,
Brower:1978wm,Zhitnitsky:1985um,Li:1986gf,Li:1987hx,Lenz:1991sa,
Burkardt:1995eb,Zhitnitsky:1995qa,Kalashnikova:1999wt,
Kalashnikova:2001df,Burkardt:2000ez,
Burkardt:2002yf,Grinstein:2008wm,Fateev:2009jf,Sazdjian:2010ku,
Ziyatdinov:2010vg,Jia:2017uul,Ma:2021yqx,Ambrosino:2023dik,
Kochergin:2024quv,Litvinov:2024riz}.  
\par
The aim of this paper is to investigate, in the large-$N_c^{}$ limit
of two-dimensional QCD, the properties of the theory concerning the
four-body sector, generated by two quark and two antiquark fields
and the possible emergence of tetraquark bound states or resonances.
As is known from the works of Witten and Coleman
\cite{Witten:1979kh,Witten:1979pi,Coleman:1985rnk}, because of
cluster factorization, multiquark states could not exist in the
large-$N_c^{}$ limit. This severe theoretical prediction was,
however, amended at a later stage by Weinberg
\cite{Weinberg:2013cfa}, who argued that even if multiquark
states would not appear in leading-order terms of $N_c^{}$,
they might still produce poles in Green's functions and
scattering amplitudes at nonleading orders of $N_c^{}$. This would
have also consequences for the corresponding couplings and decay
widths of possibly existing such states. Investigations along
this line of approach have been undertaken by several authors
\cite{Knecht:2013yqa,Cohen:2014tga,Maiani:2016hxw,
Maiani:2018pef,Lucha:2017mof,Lucha:2018dzq,Lucha:2020vgf,
Lucha:2021mwx}.
\par
Our analysis is done in the light-cone gauge, following 't~Hooft's
approach \cite{'tHooft:1973jz,'tHooft:1974hx,Callan:1975ps,
Einhorn:1976uz}. Concentrating on the case of four different quark
flavors, we derive the four-particle Green's function equations in
the color-singlet sector and the corresponding integral equations
satisfied by the scattering amplitudes. These are classified as
belonging to the ``direct'' or to the ``recombination'' channels,
according to whether the final and initial states have the same
color combinations with respect to the quark flavors or whether they
have undergone quark exchanges.
\par
One of the main challenges in dealing with confining theories is
the demonstration that the theory at hand is infrared finite.
Two-dimensional QCD offers the possibility of explicitly checking
that issue. To this aim, we introduce in the gluon propagator
an infrared regulator cutoff parameter, playing the role of a mass
term, which, at the end of the calculations is taken to zero.
\par
The infrared finiteness of the theory for quark-antiquark
color-singlet systems has been shown in \cite{'tHooft:1973jz}.
A similar property has been shown at order $1/N_c^{}$ in
\cite{Callan:1975ps}, concerning meson-meson scattering amplitudes.
In the present work, with four different quark flavors, the
recombination-type meson scattering amplitudes are of order 
$1/N_c^{}+O(1/N_c^3)$ and the derivation of their finiteness is
established in a similar way as in \cite{Callan:1975ps}. We 
show that, in the infrared limit, the scattering amplitudes
reduce to finite effective four-meson contact terms, whose
expressions are calculable in terms of overlapping integrals
involving the meson wave functions. The direct-type meson
scattering amplitudes are of order $1/N_c^2+O(1/N_c^4)$. They
are plagued by several dozens of infrared diverging terms,
involving up to four gluon propagators exchanged between different
meson clusters. These can be grouped into several categories in
which the cancelation properties of the divergences are
more easily shown. We demonstrate that each category is globally
free of divergences and the scattering amplitudes reduce to
the unitarity corrective terms arising from the recombination
contact term and a new four-meson contact term.
\par
While the demonstration of the finiteness of the scattering
amplitudes beyond the order $1/N_c^2$ seems out of reach, we
conjecture that the natural emergence of the unitarity correction
in the direct channel is the sign of a global unitarity property
satisfied by the on-mass shell scattering amplitudes. This 
leads us to complete, by a unitarization operation, the results
obtained so far and to provide finite unitarized scattering
amplitudes.
\par
Finally, the continuation of the scattering amplitudes in the
total mass-squared variable below the two-meson thresholds
allows us to obtain a bound state
equation for possibly existing tetraquark states. It is shown
that the bound state equation has generally one solution.
The properties of the latter are sensitive, however, to the
values of the effective contact terms and a detailed study
of their properties is left for a separate work.
\par
The plan of the paper is the following. In Sec. \rf{s2}, we
introduce definitions and review the properties of the
quark-antiquark systems. In Sec. \rf{s3}, we consider diquark
systems and display the differences that emerge, in the infrared
region, with respect to the color-singlet quark-antiquark case.
In Sec. \rf{s4}, we consider two-quark--two-antiquark systems and
derive the integral equations of the corresponding Green's functions
and the related meson scattering amplitudes. Section \rf{s5} is
devoted to the demonstration of the finiteness of the scattering
amplitudes up to order $1/N_c^2$. The problem of the unitarization
of the finite scattering amplitudes is considered in Sec. \rf{s6}.
Section \rf{s7} provides the bound state equation resulting from the
continuation of the scattering amplitudes below the two-meson
thresholds. In Sec. \rf{s8}, we summarize our results and
mention open questions to be resolved.
Two Appendixes provide details about the regularization scheme and
the technical operations related to the cancelation of infrared
divergences.
\par

\section{Quark-antiquark systems} \lb{s2}

We briefly sketch in this section the main results obtained so far
for quark-antiquark systems in the large-$N_c^{}$ limit
\cite{'tHooft:1974hx,Callan:1975ps,Einhorn:1976uz}, the quark fields
belonging to the fundamental representation of the color-gauge group
$SU(N_c^{})$.
They will be useful for the treatment of the more general case
of two-quark--two-antiquark systems.
\par

\subsection{Definitions and conventions} \lb{s21}

The definitions and color dependences of propagators and vertices
are presented in Fig. \rf{f1}. Small Latin letters are used for
color indices in the fundamental representation, while capital Latin
letters refer to the adjoint representation. Greek letters are
related to the Dirac spinorial indices. $T^A$ are the
color-gauge-group generators in the fundamental representation. 
\bfg
\vspace{0.5 cm}
\parbox{3 cm}
{\includegraphics[scale=0.7]{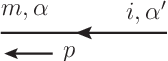}}
\hfill
\parbox{12 cm}
{
\[
\hspace{-3 cm}  
\mathrm{quark\ propagator}\ \ \ \ S_{\ \ i,\alpha\alpha'}^{m}(p)=
\delta_{\ \ i}^{m}S_{\alpha\alpha'}^{}(p),
\]
}
\par
\vspace{0.5 cm}
\parbox{3 cm}
{\includegraphics[scale=0.7]{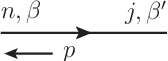}}
\hfill
\parbox{12 cm}
{
\[
\hspace{-3 cm}  
\mathrm{antiquark\ propagator}\ \ \ \ S_{\ n,\beta'\beta}^{j}(-p)=
\delta_{\ n}^{j}S_{\beta'\beta}^{}(-p),
\]
}
\par
\vspace{0.5 cm}
\parbox{3 cm}
{\includegraphics[scale=0.7]{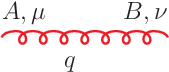}}
\hfill
\parbox{12 cm}
{
\[
\hspace{-3 cm}
\mathrm{gluon\ propagator}\ \ \ \ \ \ \ \
\delta_{AB}^{}D_{\mu\nu}(q),
\]
}
\par
\vspace{0.5 cm}
\parbox{4 cm}
{\includegraphics[scale=0.7]{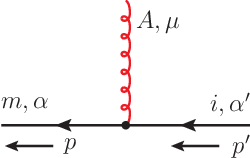}}
\hfill
\parbox{10 cm}
{
\[
\hspace{-5 cm}
\mathrm{vertex}\ \ \ \ \ \ \ \  
ig\ (T^A)_{\ \ i}^{m}\ (\gamma_{\mu}^{})_{\alpha\alpha'}^{},
\]
}
\par
\vspace{0.5 cm}
\parbox{4 cm}
{\includegraphics[scale=0.7]{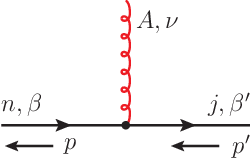}}
\hfill
\parbox{10 cm}
{
\[
\hspace{-5 cm}
\mathrm{vertex}\ \ \ \ \ \ \ \ 
ig\ (T^A)_{\ n}^{j}\ (\gamma_{\nu}^{})_{\beta'\beta}^{}.
\]
}
\vspace{0.5 cm}
\caption{Color dependences of the quark, antiquark, and gluon
propagators, and of the corresponding vertices. Small Latin letters
refer to the color indices of the fundamental representation and
capital Latin letters to the adjoint representation. Greek letters
are related to the Dirac spinorial indices. $T^A$ are the
color-gauge-group generators.} 
\lb{f1}
\efg
\par
The color-gauge-group generators $T^A$ satisfy, among others, the
following properties:
\be \lb{2e1}
(T^A)_{\ \ i}^{m}\ (T^A)_{\ n}^{j}=\frac{1}{2}(\delta_{\ \ n}^{m}
\delta_{\ i}^{j}-\frac{1}{N_c^{}}\delta_{\ \ i}^{m}\delta_{\ n}^{j}),
\ \ \ \ \ \ \ \mathrm{tr}(T^AT^B)=\frac{1}{2}\delta_{AB}.
\ee
\par
The string tension is defined through the relation\footnote{Notice
that a difference of a factor $\sqrt{2}$ exists in our definition of
the coupling constant $g$ and that of Refs.
\cite{'tHooft:1974hx,Callan:1975ps,Einhorn:1976uz}; this is due to
the difference in the definitions of the gluon fields; in Refs.
\cite{'tHooft:1974hx,Callan:1975ps,Einhorn:1976uz} the gluon fields
are defined with the double-index notation, while here, we use the
more conventional single-index notation. The relationship between
the two definitions is given by the formula 
$(T^BA_{\ \mu}^B)_{\ b}^a=A_{\ b,\mu}^a/\sqrt{2}$.}
\be \lb{2e2}
\sigma\equiv \frac{g^2N_c^{}}{4}(1-\frac{1}{N_c^2}).
\ee
\par
The light-cone components and matrices are defined as follows:
\bea \lb{2e3}
& &x^{\pm}=\frac{1}{\sqrt{2}}(x^0\pm x^1),\ \ \ \ \ x_{\pm}^{}=x^{\mp},
\nonumber \\
& &p_{\pm}^{}=\frac{1}{\sqrt{2}}(p_0^{}\pm p_1^{}),\ \ \ \ \
A_{\pm}^{}=\frac{1}{\sqrt{2}}(A_0^{}\pm A_1^{}),\nonumber \\
& &\gamma^{\pm}=\frac{1}{\sqrt{2}}(\gamma^{0}\pm \gamma^{1}),
\ \ \ \ \ \ \gamma_{\pm}^{}=\gamma^{\mp},\nonumber \\
& &(\gamma_{\pm}^{})^2=0,\ \ \ \ \
(\gamma_+^{}\gamma_-^{}+\gamma_-^{}\gamma_+^{})=2.
\eea
\par
One may also introduce two unit lightlike vectors $n$ and $\bar{n}$,
such that
\be \lb{2e4}
n^2={\bar{n}}^2=0,\ \ \ \ \ n.\bar{n}=1,
\ee
and for any vectors $V$ and $W$ one has the decompositions
\bea \lb{2e5}
& &V_{\mu}^{}=V_-^{}\bar{n}_{\mu}^{}+V_+^{}n_{\mu}^{},\ \ \ \ \
W_{\mu}^{}=W_-^{}\bar{n}_{\mu}^{}+W_+^{}n_{\mu}^{},\nonumber \\
& &n.V=V_-^{},\ \ \ \ \bar{n}.V=V_+^{},\ \ \ \ \
V.W=V_-^{}W_+^{}+V_+^{}W_-^{}.
\eea
\par
The light-cone gauge is specified with the gauge fixing condition
\be \lb{2e6}
n.A^B=A_-^B=0, \ \ \ \ \ B=1,2,\ldots,N_c^2-1.
\ee
The vertex part $A^{\mu}\gamma_{\mu}^{}$ (Fig. \rf{f1})
then reduces to $A_+^{}\gamma_{-}^{}$.
In this gauge, as well as in noncovariant gauges, ghost fields are
absent \cite{Leibbrandt:1987qv}. Furthermore, in two dimensions,
in the light-cone gauge, cubic and quartic gluon couplings are also
absent. The gluon interacts only with quarks.
\par

\subsection{Gluon and quark propagators} \lb{s22}

The gluon propagator, in the light-cone gauge (in any dimension), is
\cite{Leibbrandt:1987qv}
\be \lb{2e7}
D_{\mu\nu}^{AB}(q)=\frac{-i\delta_{AB}^{}}{(q^2+i\epsilon)}
\Big(g_{\mu\nu}^{}-\frac{(n_{\mu}^{}q_{\nu}^{}+n_{\nu}^{}q_{\mu}^{})}
{n.q}\Big)=\delta_{AB}^{}D_{\mu\nu}^{}(q).
\ee
In two dimensions, it becomes
\be \lb{2e8}
D_{\mu\nu}^{}(q)=n_{\mu}^{}n_{\nu}^{}D_{++}^{}(q),\ \ \ \ \
D_{++}^{}(q)=\frac{i}{q_-^2}.
\ee
\par
The full quark propagator $S$ is calculated with the inclusion of
the self-energy $\Sigma$, which, being a planar diagram, contributes
at leading order in $N_c^{}$.
Defining
\be \lb{2e9}
S_{\ i}^{m}=\delta_{\ i}^{m}S,\ \ \ \ \ \Sigma_{\ i}^{m}=
\delta_{\ i}^{m}\Sigma,
\ee
one has
\be
\lb{2e10}
S(p)=\frac{i}{\gamma.p-m-\Sigma+i\epsilon}.
\ee
$\Sigma$ is calculated from the Feynman diagram of Fig. \rf{f2},
with the full quark propagator in it. It is the only leading
diagram at large $N_c^{}$.
\par
\bfg 
\vspace*{1 cm}
\bc
\includegraphics[scale=1.]{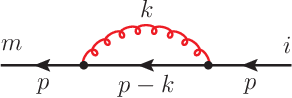}
\caption{Quark self-energy.}
\lb{f2} 
\ec
\efg
\par
Decomposing $S$ and $\Sigma$ along the $\gamma$ matrices,
\be \lb{2e11}
S=S_-^{}\gamma_+^{}+S_+^{}\gamma_-^{}+S_0^{}\ 1,\ \ \ \ \ \ \
\Sigma=\Sigma_-^{}\gamma_+^{}+\Sigma_+^{}\gamma_-^{}+\Sigma_0^{}\ 1,
\ee
one finds that $\Sigma$ is proportional to the matrix $\gamma_-^{}$
and hence $\Sigma_-^{}=\Sigma_0^{}=0$:
\be \lb{2e12}
\Sigma(p)=\gamma_-\Sigma_+^{}(p).
\ee
Furthermore, the loop integration with respect to $k_+^{}$ removes
the possible dependence of $\Sigma$ upon $p_+^{}$; one finds
\be \lb{2e13}
\Sigma_+(p_-^{})=\sigma\int\frac{dk_-^{}}{2\pi}\
\frac{\varepsilon(p_-^{}+k_-^{})}{k_-^2},
\ee
where $\varepsilon(p_-^{})=\mathrm{sgn}(p_-^{})$.
\par
The last integral is infrared divergent. 't~Hooft 
\cite{'tHooft:1974hx} has introduced an infrared cutoff
$\lambda>0$ to regulate the $k_-^{}$ integral.
We shall rather introduce a small mass term $\lambda$ in the
gluon propagator to effect the regulation. It turns out that this
procedure brings a better control of the infrared limit when we are
in the presence of multiple integrals involving several gluon
propagators in convolution with quark propagators
(cf. Appendix \rf{sa1}). Generically, we make the substitution
\be \lb{2e14}
\frac{1}{k_-^2}\ \longrightarrow\ \frac{1}{k_-^2+\lambda^2}.
\ee
We shall, however, replace $\lambda$ in the following by another
quantity, $\Lambda$, which is more suitably incorporated in the
various formulas,
\be \lb{2e15}
\Lambda\equiv \frac{\sigma}{\lambda}.
\ee
(The limit $\lambda\rightarrow 0$ is now transcribed into the
limit $\Lambda \rightarrow \infty$.)
One finally finds
\be \lb{2e16}
\Sigma_+(p_-^{})=\frac{1}{2}\varepsilon(p_-^{})
(\Lambda-\frac{2\sigma}{\pi |p_-^{}|}).
\ee
\par
The quark propagator is then
\be \lb{2e17}
S(p)=\frac{i\big\{\gamma_+^{}p_-^{}+\gamma_-^{}\big[p_+^{}-
\varepsilon(p_-^{})(\frac{\Lambda}{2}-\frac{\sigma}{\pi|p_-^{}|})
\big]+m\big\}}
{2p_+^{}p_-^{}-(|p_-^{}|\Lambda-2\sigma/\pi)-m^2+i\epsilon}.
\ee
\par
In the dynamical equations that follow, because of the particular
structure of the vertices in the light-cone gauge, where only the
$\gamma_-^{}$ matrix appears, it is the component $S_-$ of $S$ that
plays a fundamental role. We display more explicitly its expression,
\be \lb{2e18}
S_-^{}(p)=\frac{ip_-^{}}{\Big[2p_+^{}p_-^{}-|p_-^{}|\Lambda
-m^{\prime 2}+i\epsilon\Big]},
\ee
where we have defined
\be \lb{2e19}
m^{\prime 2}\equiv m^2-\frac{2\sigma}{\pi}.
\ee
\par
The quantity $|p_-^{}|\Lambda$ is a scalar and contributes to the
mass renormalization. When $\Lambda \rightarrow \infty$, the
renormalized mass tends to $\infty$, signifying that the quark,
which corresponds to a colored field, becomes unobservable in that
limit. This has occurred without making any primary hypothesis about
the physical outcome of the theory. It is then crucial to check
whether the physically observable quantities, which are expected to
be generated by color-singlet operators, do remain finite in the above
limit. Here appears one of the main advantages of using the infrared
cutoff regularization procedure, which provides us with an explicit
criterion for the distinction of observable quantities. In other
regularization schemes, such as those using the principal value
definition, one must supplement them with the requirement of gauge
invariance of the related quantities \cite{Einhorn:1976uz}.
\par

\subsection{Bound-state equation} \lb{s23}

We consider two quark fields with different flavors and free masses
$m_1^{}$ and $m_2^{}$, and specialize to the transition amplitude
where the ingoing and outgoing particles are made of the
quark-antiquark system. The quark and the antiquark are referred
by the indices 1 and $\bar 2$, respectively.
\par
We designate by $G_{1\bar 2}^{}$ the corresponding two-body Green's
function, by $G_{1\bar 2,0}^{}$ its free part and by $K_{1\bar 2}^{}$
the kernel of the corresponding integral equation; at leading order
of large $N_c^{}$, the latter reduces to the one-gluon exchange term.
\par
The integral equation has the form\footnote{In some formulas, when
too many indices are present, we write the color indices of the quarks
and of the antiquarks on the same line.}  
\bea \lb{2e20}
& &G_{1\bar 2;\alpha\beta,\beta'\alpha'}^{mn,ji}(r;p,p')=
G_{1\bar 2,0;\alpha\beta,\beta'\alpha'}^{mn,ji}(r;p)
(2\pi)^2\delta^2(p-p')\nonumber \\
& &\ \ \ \ \
+\ G_{1\bar 2,0;\alpha\beta,\xi\sigma}^{mn,sr}(r;p)\times
\int\frac{d^2p''}{(2\pi)^2}
K_{1\bar 2;\sigma\xi,\zeta\eta}^{rs,ut}(p''-p)
G_{1\bar 2;\eta\zeta,\beta'\alpha'}^{tu,ji}(r;p'',p'),
\eea
where
\be \lb{2e21}
r=\mathrm{total\ momentum},\ \ \ p=\mathrm{outgoing\ quark\
momentum},\ \ \ p'=\mathrm{ingoing\ quark\ momentum},
\ee
\bea \lb{2e22}
G_{1\bar 2,0;\alpha\beta,\beta'\alpha'}^{mn,ji}(r;p)&=&
S_{1;\alpha\alpha'}^{mi}(p)\
S_{2;\beta'\beta}^{jn}(p-r)
=\delta_{mi}^{}S_{1;\alpha\alpha'}^{}(p)
\ \delta_{jn}^{}S_{2;\beta'\beta}^{}(p-r)\nonumber \\
&=&\delta_{mi}^{}\delta_{jn}^{}
G_{1\bar 2,0;\alpha\beta,\beta'\alpha'}^{}(r;p),
\eea
$S_1^{}$ and $S_2^{}$ being the full quark propagators;
it is represented in Fig. \rf{f3}.
\bfg 
\vspace*{1 cm}
\bc
\includegraphics[scale=0.8]{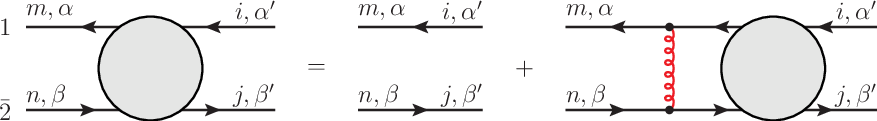}
\caption{Integral equation of the quark-antiquark Green's function.}
\lb{f3} 
\ec
\efg
\par
The expression of $K_{1\bar 2}^{}$ is
\bea \lb{2e23}
K_{1\bar 2;\sigma\xi,\zeta\eta}^{rs,ut}(p''-p)&=&
-g^2(T^A)_{rt}^{}(T^A)_{us}^{}(\gamma^{\mu})_{\sigma\eta}^{}
(\gamma^{\nu})_{\zeta\xi}^{}D_{\mu\nu}^{}(p''-p)\nonumber \\
&=&-\frac{g^2}{2}(\delta_{rs}^{}\delta_{ut}^{}
-\frac{1}{N_c^{}}\delta_{rt}^{}\delta_{us}^{})
(\gamma_-^{})_{\sigma\eta}^{}
(\gamma_-^{})_{\zeta\xi}^{}D_{++}^{}(p''-p).
\eea
\par
The Green's functions $G$ and $G_0^{}$ can be decomposed along the
basis of the tensor products of the two sets of independent $\gamma$
matrices ($\gamma_+^{},\gamma_-^{},\gamma_-^{}\gamma_+^{},
\gamma_+^{}\gamma_-^{}$)
related to the quark 1 and the antiquark $\bar 2$, respectively.
Because of the particular structure of the
kernel $K$, which is proportional to two $\gamma_-$ matrices,
each acting on the quark or the antiquark lines, and the fact that
$\gamma_{\pm}^2=0$, it is only the component $G_{-,-}^{}$
of $G$ accompanying the matrices $\gamma_+^{}\times \gamma_+^{}$ that
contributes to the dynamical part of the equation, the other
components being determined from it by kinematic relations.
Similarly, it is the components $S_{1,-}$ and $S_{2,-}$ of $S_1$
and $S_2$, respectively, that participate in the above
equation. It is therefore sufficient to project the integral
equation on the above sector, simplifying it from redundant indices
and concentrating on the color indices.
On the other hand, the quark propagators that surround a gluon propagator
at its vertices produce by contraction of the $\gamma_+$ matrices a
factor 2 at each vertex \cite{'tHooft:1974hx}, and hence a factor 4
for the gluon propagator contribution in the sector of the
$\gamma_+^{}\times \gamma_+^{}$ matrices.
To simplify notations, one can remove from the various quantities the
corresponding $\gamma$-matrix indices and adopt the following
redefinitions:
\bea \lb{2e24}
& &S_{a,-}\ \rightarrow\ S_a,\ \ \ a=1,2,\ \ \  
D_{++}\ \rightarrow\ D=\frac{i}{q_-^2+\lambda^2}, \ \ \
G_{-,-}\ \rightarrow\ G,\ \ \ 
\ \ S_{1,-}S_{2,-}=G_{0,-,-}\ \rightarrow\ G_0.\nonumber \\
& &
\eea
Removing the system indices $1\bar 2$ and the integration symbols,
the integral equation takes the following compact form:

\be \lb{2e25}
G_{\ n,\ i}^{m,\ j}=G_{0,\ n,\ i}^{\ m,\ j}
+\,G_{0,\ n,\ r}^{\ m,\ s}\ 4\,K_{\ s,\ t}^{r,\ u}\
G_{\ u,\ i}^{\,t,\ j},
\ee
with
\bea \lb{2e26}
& &G_{0,\ n,\ i}^{\ m,\ j}=\delta_{\ \,i}^m\delta_{\ n}^jG_0,\ \ \ \ \
K_{\ s,\ t}^{r,\ u}=\Big(-\frac{g^2}{2}D\Big)
(\delta_{\ s}^r\delta_{\ t}^u
-\frac{1}{N_c}\delta_{\ t}^r\delta_{\ s}^u),\nonumber \\
& &\Big(G_0K\Big)_{\ n,\ t}^{m,\,u}=G_0K\frac{1}{N_c^{}}
\Big[\delta_{\ n}^m\delta_{\ t}^u-\frac{N_c{}}{(N_c^{2}-1)}
(\delta_{\ t}^m\delta_{\ n}^u
-\frac{1}{N_c}\delta_{\ n}^m\delta_{\ t}^u)\Big],\ \ \ \
K=(-2\sigma D).
\eea
\par
Since the quark and antiquark fields belong to the fundamental and
antifundamental representations of the color-gauge group $SU(N_c)$,
the quark-antiquark system should belong either to the singlet or
to the adjoint representation. Global color being preserved by the
interaction, this feature is also exhibited on the decomposition
of the quantity $(G_0^{}K)$ of Eqs. (\rf{2e26}) and transmitted to
the intermediate states that saturate the Green's function. 
The latter can also be decomposed along the above representations.
We therefore introduce a similar decomposition of $G$,
\be \lb{2e27}
G_{\ n,\ i}^{m,\ j}=\delta_{\ n}^m\delta_{\ i}^j\, G_s
+(\delta_{\ i}^m\delta_{\ n}^j
-\frac{1}{N_c}\delta_{\ n}^m\delta_{\ i}^j)\, G_{adj},
\ee
where $G_s$ and $G_{adj}$ represent the parts of the Green's function
that contain intermediate states belonging to the singlet and adjoint
representations, respectively. The integral equation (\rf{2e25})
then splits into two decoupled independent equations,
\bea
\lb{2e28}
& &G_s=\frac{1}{N_c}G_0\,+\,G_0(4K)G_s,\\
\lb{2e29}
& &G_{adj}=G_0-\frac{1}{(N_c^2-1)}G_0(4K)G_{adj}.
\eea
It turns out, as will be shown in the following, that
the kernel part, $G_0(4K)$, with a positive sign corresponds to an
attractive interaction and may therefore lead to the existence of
bound states.
This is the case of $G_s$. On the other hand, the quark-antiquark
system in its adjoint representation, with a negative sign in front of
$G_0(4K)$, is submitted to an internal repulsive interaction and could
not lead to the appearance of bound states.
To be observable, unbound states should display properties close to
those of scattering states. However, we have seen in Sec. \rf{s22},
under the effect of the infrared singularity, individual quark states
escape observational criteria. Also note that the kernel in the adjoint
representation is damped by a factor of $1/(N_c^2-1)$. In the
large-$N_c$ limit, it tends to zero. We shall concentrate in
the following on the color-singlet sector of the quark-antiquark
system.
\par
In Eq. (\rf{2e28}), the factor $1/N_c$ in front of $G_0$ fixes the
normalization constant of the possibly existing bound state wave
functions. Otherwise, it does not have any influence on the dynamics of
the system. This is easily seen by redefining $G_s$ with a similar
coefficient (dropping the subscript $s$ from the redefined function),
\be \lb{2e30}
G_s=G/N_c \ \ \ \Longrightarrow \ \ \ \ G=G_0\,+\,G_0(4K)G.
\ee
\par
Assuming the existence of bound states with wave functions $\phi$,
their equation is determined by the homogeneous part of Eq.
(\rf{2e30}), which we display in explict form,
\bea \lb{2e31}
\phi(r,p)&=&-\frac{p_-}
    {\big[p^2-|p_-^{}|\Lambda
        -m_1^{\prime 2}+i\epsilon\big]}
\times \frac{(r_-^{}-p_-^{})}
    {\big[(r-p)^2-|r_-^{}-p_-^{}|\Lambda
        -m_2^{\prime 2}+i\epsilon\big]}\nonumber \\
& & \times (8\sigma)
\int\frac{d^2p''}{(2\pi)^2}
\frac{i}{(p_-''-p_-^{})^2+\lambda^2}\phi(r,p''),
\eea
$\Lambda$ and $m^{\prime 2}$ being defined in Eqs. (\rf{2e15}) and
(\rf{2e19}).
The Dirac matrix structure of the total wave function can be
reconstructed by relating it, together with its adjoint,
to the tensor basis of the $\gamma$ matrices of the outgoing and
ingoing sectors, respectively. To this end, it is sufficient to
use the relationship
$(\gamma_+)_{\alpha\alpha'}^{}(\gamma_+)_{\beta'\beta}^{}=
(\gamma_+)_{\alpha\beta}^{}(\gamma_+)_{\beta'\alpha'}^{}$. In this
basis, $\phi$ corresponds to the component $\phi_-^{}$.
\par
The interaction being lightlike instantaneous, the integration with
respect to $p_+''$ can be done, removing $p_+^{}$ from the
integral. Defining the integrated wave function as
\be \lb{2e32}
\varphi(r,p_-^{})=\int \frac{dp_+^{}}{2\pi}\ \phi(r,p),
\ee
which represents the lightlike instantaneous limit of $\phi$, one can
integrate Eq. (\rf{2e31}) with respect to $p_+^{}$ on both sides.
It concerns on the right-hand side the external propagators and gives
a nonzero result only if the conditions $p_-^{}>0$ and
$p_-^{}<r_-^{}$ are satisfied ($r_-^{}$ and $r_+^{}$ being both
positive), which are then reflected in the wave function of the
left-hand side [cf. Eq. (\rf{a2e2}) of Appendix \rf{sa2}].
We hence assume that the wave function $\varphi(r,p_-^{})$
satisfies those conditions,
\be \lb{2e33}
\varphi(r,p_-^{})=\varphi(r,p_-^{})\theta(p_-^{}(r_-^{}-p_-^{})).
\ee
One obtains
\be \lb{2e34}
\Big[\ r_+^{}-\Lambda
  -\frac{m_1^{\prime 2}}{2p_-^{}}
  -\frac{m_2^{\prime 2}}{2(r_-^{}-p_-^{})}\ \Big]\
\varphi(r,p_-^{})=-2\sigma
\int_0^{r_-}\frac{dp_-''}{(2\pi)}\frac{1}{(p_-''-p_-^{})^2+\lambda^2}\
\varphi(r,p_-'').
\ee
The integral in Eq. (\rf{2e34}) is divegent in the limit
$\lambda\rightarrow 0$. The details of the separation of the
divergent part are presented in Eqs. (\rf{a1e1})-(\rf{a1e4}) of
Appendix \rf{sa1}. One obtains for the right-hand side of Eq.
(\rf{2e34}) the expression
\be \lb{2e35}
-2\sigma\int_0^{r_-}\frac{dp_-''}{(2\pi)}
\frac{1}{(p_-''-p_-^{})^2}\,(\varphi(r,p_-'')-\varphi(r,p_-))
-\Big(\Lambda-\frac{2\sigma}{\pi}
(\frac{1}{2p_-}+\frac{1}{2(r_--p_-)})\Big)\,\varphi(r,p_-).
\ee
The divergent part cancels the factor $\Lambda$
of the left-hand side, while the last two finite contributions cancel 
the finite renormalization parts contained in $m_1^{\prime 2}$ and
$m_2^{\prime 2}$, bringing them to their initial values $m_1^2$ and
$m_2^2$. The equation becomes
\bea \lb{2e36}
& &\Big[\ r_+^{}
-\frac{m_1^{2}}{2p_-^{}}
-\frac{m_2^{2}}{2(r_-^{}-p_-^{})}\ \Big]\
\varphi(r,p_-^{})=-2\sigma
\int_0^{r_-}\frac{dp_-''}{(2\pi)}\frac{1}{(p_-''-p_-^{})^2}\
(\varphi(r,p_-'')-\varphi(r,p_-^{})).\nonumber \\
& &
\eea
Introducing new variables $x$ and $y$,
\be \lb{2e37}
x=\frac{p_-^{}}{r_-^{}},\ \ \ \ \ 0\le x\le 1,
\ \ \ \ \ \ y=\frac{p_-''}{r_-^{}},\ \ \ \ \ 0\le y\le 1,
\ee
the equation takes the form
\be \lb{2e38}
\Big[\ r^2-\frac{m_1^2}{x}-\frac{m_2^2}{(1-x)}\ \Big]\
\varphi(x)=-(\frac{2\sigma}{\pi}) \int_0^{1}dy
\frac{(\varphi(y)-\varphi(x))}{(y-x)^2}.
\ee
This is the 't~Hooft equation. It may take slightly different forms,
according to the way of grouping the regularized terms.
\par
't~Hooft has shown that the mass-squared operator is positive
definite and self-adjoint in the space of functions $\varphi(x)$
that vanish at the boundaries $x=0$ and $x=1$. The spectrum of
eigenvalues is discrete, characterized with an integer quantum
number $n$ $(=0,1,2,\ldots)$, and corresponds to a confining
potential. The eigenfunctions form a complete set and are
orthogonal among themselves for different quantum numbers.
\par
For large values of $n$, the eigenfunctions can be approximated
by
\be \lb{2e39}
\varphi_n^{}(x)\sim \sin(n\pi x),
\ee
and the spectrum is approximately
\be \lb{2e40}
r_n^2\simeq 2\sigma\pi n+
(m_1^{\prime 2}+m_2^{\prime 2})\ln(n)+\mathrm{const}.
\ee
The leading term in $n$ justifies the identification of $\sigma$ with
the string tension \cite{Ida:1977uy}.
When the quark masses tend to zero, the ground state mass also tends 
to zero, with $\varphi$ tending to a constant.
\par
One important property of Eq. (\rf{2e38}) is the disappearance of
the cutoff factor $\Lambda$; this is a consistency check of the
present formalism for the evaluation of all observable quantities.
\par
The complete wave function $\phi$ can be determined from Eq.
(\rf{2e31}), using in its right-hand side Eqs. (\rf{2e32}) and
(\rf{2e34}). One finds
\be \lb{2e41}
\phi(r,p)=4iG_0\Big[\ r_+^{}-\Lambda
-\frac{m_1^{\prime 2}}{2p_-^{}}
-\frac{m_2^{\prime 2}}{2(r_-^{}-p_-^{})}\ \Big]
\varphi(r,p_-^{}).
\ee
\par
The normalization condition of the wave function $\phi$ according
to the Bethe--Salpeter equation, after restoration of the $\gamma$
matrices and of the factor $1/N_c$ of Eq. (\rf{2e28}), imposes that
of the reduced wave functions $\varphi$ in the limit
$\Lambda \rightarrow \infty$,
\be \lb{2e42}
\int_0^1dx \varphi_m^*(x)\varphi_n(x)=
\delta_{mn}\frac{\pi}{N_c}.
\ee
The same result can also be found directly from Eq. (\rf{2e30}),
after restoration of the factor $1/N_c$ in the inhomogeneous part,
with the use of the wave function (\rf{2e41}).
\par
The quark-antiquark scattering amplitude $\mathcal{T}$, which is
generated by the one-gluon-exchange diagram, is proportional,
in its $\gamma$-matrix form, to the tensor product
$\gamma_-^{}\times \gamma_-^{}$. Using, in matrix form, its defining
equation, $G=G_0^{}/N_C^{}+G_0^{}\mathcal{T}G_0^{}$, one again finds
that it is the component $G_{-,-}^{}$ of $G$, accompanying the matrices
$\gamma_+^{}\times \gamma_+^{}$, that contributes to the dynamical
part of the equation. The integral equation of the component
$\mathcal{T}_{+,+}^{}$, which accompanies the
$\gamma_-^{}\times \gamma_-^{}$ matrices of $\mathcal{T}$, takes
the following form for the color-singlet sector (omitting
the indices),
\be \lb{2e43}
\mathcal{T}=\frac{1}{N_c^{}}K+4KG_0^{}\mathcal{T},
\ee
or more explicitly
\be \lb{2e44}
\mathcal{T}(r;p,p')=-\frac{2\sigma}{N_c} D(p-p')
-8\sigma \int\frac{d^2k}{(2\pi)^2}
D(k-p)\ S_{1}(k)S_{2}(k-r)\
\mathcal{T}(r;k,p').
\ee
This equation has been solved in
\cite{Callan:1975ps}.\footnote{In that reference, the multiplicative
factor 4 of the kernel has been absorbed in a redefinition of the
quark propagators, with a multiplicative factor 2 for each.}
The solution is
\be \lb{2e45}
\mathcal{T}(r;x,x')=-\frac{2\sigma}{N_c} \frac{i}{r_-^2(x-x')^2}
+\frac{i}{N_c^{}}\sum_n^{}\frac{\widetilde{\phi}_n^{}(r,x)
\widetilde{\phi}_n^*(r,x')}{(r^2-r_n^2)}.
\ee
$\widetilde{\phi}_n^{}(r,x)$ is related to the eigenfuction
$\varphi_n^{}(x)$ through the relation
\be \lb{2e46}
\widetilde{\phi}_n^{}(r,x)=-\frac{i}{r_-^{}}
(\frac{2\sigma}{\sqrt{\pi}})
\int_0^1dy\frac{\varphi_n^{}(y)}{(y-x)^2},
\ee
where now the canonical orthonormality condition
$\int_0^1dx \varphi_m^*(x)\varphi_n(x)=\delta_{mn}^{}$
has been used for the $\varphi_n^{}$s to make the
$1/N_c$ dependence of $\mathcal{T}$ explicit.
\par
For $x<0$ or $x>1$, the right-hand side of Eq. (\rf{2e46}) is a
regular function of $x$ and allows the analytic continuation of
$\widetilde{\phi}_n^{}(r,x)$ to that domain and hence of
$\mathcal{T}(r;x,x')$ when it is considered in unphysical domains
of the momenta. Using Eq. (\rf{2e38}), which also allows the analytic
continuation of $\varphi_n^{}(x)$ to the domains $x<0$ and $x>1$, the
relation (\rf{2e46}) takes, for any $x$, the form
\be  \lb{2e47}
\widetilde{\phi}_n^{}(r,x)=\frac{i\sqrt{\pi}}{r_-^{}}
\Big[\ r_n^2-2\Lambda |r_-^{}|\theta(x(1-x))
-\frac{m_1^{\prime 2}}{x}
-\frac{m_2^{\prime 2}}{(1-x)}\ \Big]\varphi_n^{}(x).
\ee
Notice that the divergent part of $\widetilde{\phi}_n^{}(r,x)$
in the limit $\Lambda\rightarrow \infty$ comes from
the contribution of the domain $0\leq x\leq 1$.  
\par
The relationship between the bound state wave functions defined
from the Green's function and those related to $\mathcal{T}$
is
\be \lb{2e48}
\phi_n^{}(r,x)=\frac{2}{\sqrt{N_c^{}}}\,G_0^{}\,
\widetilde{\phi}_n^{}(r,x).
\ee
For completeness, we also display explicitly the relation between
the Green's function and the scattering amplitude after restoration
of the $\gamma$-matrix component indices,
\be \lb{2e49}
G_{-,-}^{}=\frac{G_{0,-,-}^{}}{N_C^{}}+G_{0,-,-}^{}\,
4\mathcal{T}_{+,+}\,G_{0,-,-}^{}.
\ee
\par
One notices from Eqs. (\rf{2e45}), (\rf{2e47}) and (\rf{2e48}) that
the complete wave function $\phi$ and the scattering
amplitude $\mathcal{T}$ are cutoff dependent and therefore not directly
observable. The quark-antiquark scattering amplitude contributes,
however, to the calculation of the meson-meson scattering amplitudes.
One then has to check that the latter, on the mass-shell, are cutoff
independent. This has been verified, in simple cases of quark
flavors, at leading order $1/N_c^{}$  \cite{Callan:1975ps}. This
results from genuine compensations between the divergent terms of
the quark-antiquark scattering amplitude and the damping factors of
the quark propagators, after the loop integrations with respect to
the $+$ component of the loop momenta are done.
\par
Finally, we consider the effects of $1/N_c^{}$-order subleading
contributions into the results obtained thus far.
The gluon propagator receives radiative corrections from quark loop
insertions. Considering $N_f^{}$ quark flavors, the gluon propagator
obtains the form \cite{Callan:1975ps}
\be \lb{2e50}
D_{++}^{}(q)=\frac{i}{q_-^2+\lambda^2
+\lambda|{q_-}|N_f^{}/(\pi N_c^{})}.
\ee
It can be verified that the net effect of the modification of the
gluon propagator on the evaluation of singular contributions amounts
to the change of the infrared cutoff parameter in the form
\be \lb{2e51}
\Lambda\rightarrow \Lambda \Big(1-\frac{N_f^{}}{\pi^2N_c^{}}\Big).
\ee
This change is transmitted to the expression of the quark self-energy
(\rf{2e16}), which now becomes
\be \lb{2e52}
\Sigma_+(p_-^{})=\frac{1}{2}\varepsilon(p_-^{})
\Big(\Lambda (1-\frac{N_f^{}}{\pi^2N_c^{}})
-\frac{2\sigma}{\pi |p_-^{}|}\Big).
\ee
One notices that the finite part of $\Sigma_+$ has not been modified,
which means that the mass redefinition (\rf{2e19}) has not been
affected by the presence of the quark loops. One therefore may
redefine the infrared cutoff $\Lambda$ as represented by the
right-hand side of the formula (\rf{2e51})
and ignore the presence of quark loops at the next subleading order
in $1/N_c^{}$.
\par
Crossed diagrams of gluon propagators and vertex corrections to
the gluon-quark coupling, representing nonplanar diagrams, contribute
only to subleading orders in $1/N_c^{}$. However, in two dimensions
and in the light-cone gauge, because of the changes of sign in
front of one of the $i\epsilon$ factors of the quark propagators,
their contribution generally vanishes during loop integrations.
In more complicated diagrams, remainders may survive with coefficients
depending on the infrared cutoff $\Lambda$, such that in the infrared
limit $\lambda\rightarrow 0$, because of the spectral conditions
satisfied by the meson wave functions and due to the quark
momentum routings, they also vanish.
\par
In conclusion, when evaluating below multiquark Green's functions
or meson-meson scattering amplitudes, the presence of quark loops,
crossed diagrams and vertex corrections can be ignored at the next
subleading order in $1/N_c^{}$.
\par

\section{Diquark systems} \lb{s3}

We consider in this section diquark systems, made of two quark
fields. Each quark belonging to the fundamental representation of the
color-gauge group $SU(N_c)$, it is evident that such systems cannot
be in the color-singlet representation. Rather, they belong either
to the symmetric representation $[1,1]$ (in the terminology of Young
tableaux) or to the antisymmetric one $[2]$, corresponding, in the
case of $SU(3)$, to the representations $\mathbf{6}$  and
$\overline\mathbf{3}$, respectively.
One would expect, on the basis of a possible infrared
cutoff presence, that they could not be observed in isolation.
However, diquark systems, together with antidiquark ones, might
be embedded in larger color-singlet systems, in which case a global
cutoff cancelation might occur. This is why a separate study of
diquark systems would be of interest in its own.
\par
The integral equation satisfied by the diquark Green's function is
very similar to that of the quark-antiquark system, except for the
routings of the color and Dirac indices (cf. Fig. \rf{f4}), the
second quark being represented by the index 2.
\bfg 
\vspace*{1 cm}
\bc
\includegraphics[scale=0.8]{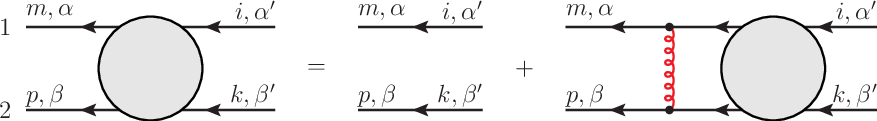}
\caption{Integral equation of the diquark Green's function.}
\lb{f4} 
\ec
\efg
\par
One has,
\be \lb{3e1}
G_{12,0;\alpha\beta,\beta'\alpha'}^{mp,ki}(r;p)=
\delta_{mi}^{}S_{1;\alpha\alpha'}^{}(p)
\ \delta_{pk}^{}S_{2;\beta\beta'}^{}(r-p),
\ee
\bea \lb{3e2}
K_{12;\sigma\xi,\zeta\eta}^{rs,ut}(p''-p)&=&
-g^2(T^A)_{rt}^{}(T^A)_{su}^{}(\gamma^{\mu})_{\sigma\eta}^{}
(\gamma^{\nu})_{\xi\zeta}^{}D_{\mu\nu}^{}(p''-p)\nonumber \\
&=&-\frac{g^2}{2}\Big[\frac{1}{2}(1-\frac{1}{N_c^{}})
    (\delta_{rt}^{}\delta_{su}^{}+\delta_{ru}^{}\delta_{st}^{})
    -\frac{1}{2}(1+\frac{1}{N_c^{}})
    (\delta_{rt}^{}\delta_{su}^{}-\delta_{ru}^{}\delta_{st}^{})\Big]
\nonumber \\
& &\ \ \ \ \ \ \ \times(\gamma_-^{})_{\sigma\eta}^{}
(\gamma_-^{})_{\xi\zeta}^{}D_{++}^{}(p''-p).
\eea
\par
Here again, the Green's functions $G$ and $G_0^{}$ can be decomposed
along the basis of the tensor products of the two sets of independent
Dirac matrices related to the quarks 1 and 2, respectively.
As in the quark-antiquark case, one finds that it is only the component
$G_{-,-}^{}$ of $G$ accompanying the matrices
$\gamma_+^{}\times \gamma_+^{}$ that contributes to the dynamical part
of the equation, together with the components $S_{1,-}$ and $S_{2,-}$
of $S_1$ and $S_2$, respectively. Projections similar to those of Eq.
(\rf{2e24}) can then be adopted.
The integral equation takes the following compact form:
\be \lb{3e3}
G_{\ \ \ k i}^{m p,}=G_{0,\ \ \ k i}^{\ m p,}
+\,G_{0,\ \ \ s r}^{\ m p,}\,4\,K_{\ \ \ u t}^{r s,}\
G_{\ \ \ k i}^{\,t u,},
\ee
with
\bea \lb{3e4}
& &G_{0,\ \ \ k i}^{\ m p,}=\delta_{\ \,i}^m\delta_{\ k}^pG_0,
\nonumber \\
& &K_{\ \ \ u t}^{r s,}=\Big(-\frac{g^2}{2}D\Big)
\Big[\frac{1}{2}(1-\frac{1}{N_c^{}})
(\delta_{\ t}^{r}\delta_{\ u}^{s}+\delta_{\ u}^{r}\delta_{\ t}^{s})
-\frac{1}{2}(1+\frac{1}{N_c^{}})
(\delta_{\ t}^{r}\delta_{\ u}^{s}-\delta_{\ u}^{r}\delta_{\ t}^{s})
\Big],\nonumber \\
& &\Big(G_0K\Big)_{\ \ \ u t}^{m p,}=G_0K
\Big[\frac{1}{2}\frac{1}{(N_c^{}+1)}
(\delta_{\ t}^{m}\delta_{\ u}^{p}+\delta_{\ u}^{m}\delta_{\ t}^{p})
-\frac{1}{2}\frac{1}{(N_c^{}-1)}
(\delta_{\ t}^{m}\delta_{\ u}^{p}-\delta_{\ u}^{m}\delta_{\ t}^{p})
\Big],\nonumber \\
& &K=(-2\sigma D).
\eea
\par
As stated earlier, the diquark sytem should belong either to
the symmetric $[1,1]$ or to the antisymmetric $[2]$ representations
of the color group $SU(N_c^{})$. This feature is also exhibited in
the decomposition of the quantity $(G_0^{}K)$ of Eqs. (\rf{3e4}).
The Green's function itself can then be decomposed along the above
representations,
\be \lb{3e5}
G_{\ \ \ k i}^{m p,}=
\frac{1}{2}
(\delta_{\ i}^{m}\delta_{\ k}^{p}+\delta_{\ k}^{m}\delta_{\ i}^{p})
\,G_S^{}
-\frac{1}{2}
(\delta_{\ i}^{m}\delta_{\ k}^{p}-\delta_{\ k}^{m}\delta_{\ i}^{p})
\,G_A^{}
\ee
where $G_S$ and $G_A$ refer to the symmetric and antisymmetric parts
of the Green's function, respectively. The integral equation (\rf{3e5})   
then splits into two decoupled independent equations,
\bea
\lb{3e6}
& &G_S^{}=G_0+\frac{4}{(N_c^{}+1)}G_0K\,G_S^{},\\
\lb{3e7}
& &G_A^{}=G_0-\frac{4}{(N_c-1)}G_0K\,G_A^{}.
\eea
Among the kernels of the two equations, it is that of $G_A^{}$ that
is attractive and could lead to the formation of a bound
state.\footnote{In comparison to Eqs. (\rf{2e28}) and (\rf{2e29}), here
the quark 2 propagator in $G_0^{}$ introduces an additional minus sign
with respect to that of the antiquark $\bar 2$.} We shall therefore
concentrate on the antisymmetric representation. Proceeding as in the
quark-antiquark case and assuming the existence of bound states with
wave functions $\phi_A^{}$, one obtains the equation
\bea \lb{3e8}
\phi_A^{}(r,p)&=&-\frac{4}{(N_c^{}-1)}\frac{p_-}
    {\big[p^2-|p_-^{}|\Lambda
        -m_1^{\prime 2}+i\epsilon\big]}
\times \frac{(r_-^{}-p_-^{})}
    {\big[(r-p)^2-|r_-^{}-p_-^{}|\Lambda
        -m_2^{\prime 2}+i\epsilon\big]}\nonumber \\
& & \times (2\sigma)
\int\frac{d^2p''}{(2\pi)^2}\frac{i}{(p_-''-p_-^{})^2+\lambda^2}
\phi_A^{}(r,p'').
\eea
Defining the lightlike instantaneous limit of $\phi_A^{}$ as in Eq.
(\rf{2e32}),
\be \lb{3e9}
\varphi_A^{}(r,p_-^{})=\int \frac{dp_+^{}}{2\pi}\ \phi_A^{}(r,p),
\ee
one arrives at the equation
\bea \lb{3e10}
& &\Big[\ r_+^{}-\Lambda
  -\frac{m_1^{\prime 2}}{2p_-^{}}
  -\frac{m_2^{\prime 2}}{2(r_-^{}-p_-^{})}\ \Big]\
\varphi_A^{}(r,p_-^{})\nonumber \\
& &\ \ \ \ \ \ \ \ \ =-\frac{2\sigma}{(N_c^{}-1)}\Big[\
\int_0^{r_-}\frac{dp_-''}{(2\pi)}\frac{1}{(p_-''-p_-^{})^2}\
(\varphi_A^{}(r,p_-'')-\varphi_A^{}(r,p_-))\ \Big]\nonumber \\
& &\ \ \ \ \ \ \ \ \ \ -\frac{1}{(N_c^{}-1)}\Big(\Lambda
-\frac{2\sigma}{\pi}(\frac{1}{2p_-}+\frac{1}{2(r_--p_-)})\Big)\,
\varphi_A(r,p_-).
\eea
We observe, as expected, that the infrared cutoff $\Lambda$ is
not canceled between both sides of the equation. This means that
the diquark sytem, considered in isolation, cannot lead to observable
bound states, since in the limit $\Lambda\rightarrow\infty$ the only
solution of the equation is $\varphi_A^{}=0$. Nevertheless, one might
consider situations where the diquark system is embedded within a
larger color-singlet system
\cite{Jaffe:1976ih,Anselmino:1992vg,Jaffe:2003sg,
Shuryak:2003zi,Maiani:2004vq,Maiani:2005pe}, in which case its
residual cutoff $\Lambda$ might be canceled by contributions coming
from the other subsystems.
What would then be the contribution of the finite
part of the diquark system, considered as if it was in isolation?
For this, it is sufficient to remove from Eq. (\rf{3e10}) its
divergent part, as well as, for simplicity, the finite renormalization
parts proportional to $(2\sigma/\pi)$, which could only affect the
values of the low-lying bound-state masses, to obtain
\bea \lb{3e11}
& &\Big[\ r_+^{}
-\frac{m_1^{2}}{2p_-^{}}
-\frac{m_2^{2}}{2(r_-^{}-p_-^{})}\ \Big]\
\varphi_A^{}(r,p_-^{})\nonumber \\
& &\ \ \ \ \ \ \ \ \ \ \ \ \ \ \ =-\frac{2\sigma}{(N_c^{}-1)}\Big[\
\int_0^{r_-}\frac{dp_-''}{(2\pi)}\frac{1}{(p_-''-p_-^{})^2}\
(\varphi_A^{}(r,p_-'')-\varphi_A^{}(r,p_-))\ \Big].
\eea
This equation has the same structure as that of the quark-antiquark
system, except that the coupling constant $\sigma$ is now replaced by
$\sigma/(N_c^{}-1)$. This means that the spectrum of the bound states
of the diquark system is very similar to that of the quark-antiquark
system with a damping of the energy gaps due to the scaling 
$\sigma\rightarrow \sigma/(N_c^{}-1)$. In particular, in the limit
$N_c^{}\rightarrow\infty$, the energy gaps between the bound states
tend to zero and the whole spectrum shrinks to the two-quark mass
threshold. In the real world, where $N_c^{}=3$, the damping factor
$1/(N_c^{}-1)$ is $1/2$.
\par

\section{Two-quark--two-antiquark systems} \lb{s4}

We consider in this section the case of two-quark--two-antiquark
systems, out of which tetra\-quarks are expected to be formed.
The quark flavors are assumed to be different from each other; this
case avoids mixing problems with ordinary meson states. The quarks
are denoted with the labels 1 and 3, with masses $m_1^{}$ and
$m_3^{}$, and the antiquarks with labels $\bar 2$ and $\bar 4$,
with masses $m_2$ and $m_4$, respectively.
\par
In relation with the present investigation and the large-$N_c^{}$
considerations, let us, at this stage, outline the following property
of multiquark states.
Contrary to the ordinary baryonic case, where large $N_c^{}$
requires increasing of the number of constituent quarks with
$N_c^{}$, the description of multiquark states, other than ordinary
baryons, can be realized by several different types of representation
(cf. Ref. \cite{Lucha:2021mwx}, Sec. 3.2).
Tetraquarks, in a diquark antisymmetric representation, can be
described in $(N_c^{}-2)$ different ways. One way corresponds to
the case where one has $(N_c^{}-1)$ external quarks linked with a
single Wilson line to $(N_c^{}-1)$ external antiquarks. Other
representations are obtained by decreasing the number of external
quarks and antiquarks, but increasing in parallel the number of
links between them. The extreme case to this corresponds to two
external quarks and two external antiquarks linked to each other
with $(N_c^{}-2)$ Wilson lines (cf. Ref. \cite{Lucha:2021mwx},
Eq. (43) and Fig. 19). This last case is the most practical
one, since it preserves the same external description of the state
as in the case $N_c^{}=3$; the variation of $N_c^{}$ affects only
the parameters related to the internal structures, in much analogy
with the ordinary meson case. We shall stick, in the present work,
to that representation of tetraquark states, also including the
symmetric diquark representation and other equivalent representations
obtained by Fierz transformations.

\subsection{Green's functions} \lb{s41}

The Green's function corresponding to the four ingoing and outgoing
particles made of the quarks 1 and 3 and the antiquarks $\bar 2$ and
$\bar 4$ is represented in Fig. \rf{f5}.
\bfg 
\vspace*{1 cm}
\bc
\includegraphics[scale=0.7]{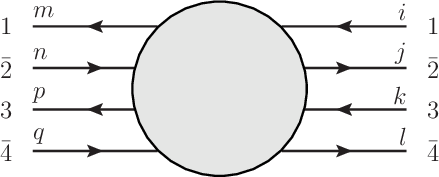}
\caption{Four-body Green's function with two quarks, labeled 1 and 3,
and two antiquarks, labeled $\bar 2$ and $\bar 4$. Latin letters
refer to color indices.}
\lb{f5} 
\ec
\efg
\par
Most of the properties, encountered with the two-body systems in
Secs. \rf{s2} and \rf{s3}, concerning $N_c$-leading powers of
diagrams and Dirac matrices, are also met in the present case.
The $N_c$-leading diagrams are produced by one-gluon exchange kernels.
Because the latter are proportional to the tensor product of two
Dirac $\gamma_-$ matrices, the final dynamical equation is governed
by the component $G_{-,-,-,-}$, which accompanies the tensor
product of four $\gamma_+$ matrices, each acting on a single quark
line; at the same time, the quark propagators of the free part of the
Green's function act in this equation with their $S_{a,-}$ component
($a=1,\ldots,4$) [Eq. (\rf{2e18})]. Notations, similar to those of Eq.
(\rf{2e24}), can then be adopted,
\bea \lb{4e1}
& &S_{a,-}\ \rightarrow\ S_a\ \ \ (a=1,2,3,4),\ \ \ \ 
D_{++}\ \rightarrow\ D, \ \ \ \ \ G_{-,-,-,-}\ \rightarrow\ G,
\nonumber \\
& &S_{1,-}S_{2,-}S_{3,-}S_{4,-}=G_{0,-,-,-,-}\ \rightarrow\ G_0,
\ \ \ \ \ 
S_{a,-}S_{b,-} \rightarrow\ G_{ab,0}\ \ \ (a,b=1,2,3,4,\ a\neq b).
\nonumber \\
\eea
\par
The contributions of the one-gluon exchange kernels are similar
to those met in the quark-antiquark and the diquark cases, Eqs.
(\rf{2e26}) and (\rf{3e4}), in which we have to include now the
corresponding particle indices,
\bea
\lb{4e2}
& &\Big(G_{ab,0}K_{ab}\Big)_{\ n,\ t}^{m,\,u}=G_{ab,0}
K_{ab}\frac{1}{N_c^{}}
\Big[\delta_{\ n}^m\delta_{\ t}^u-\frac{N_c{}}{(N_c^{2}-1)}
(\delta_{\ t}^m\delta_{\ n}^u
-\frac{1}{N_c}\delta_{\ n}^m\delta_{\ t}^u)\Big]\nonumber \\
& &\ \ \ \ \ \ \ \ \ \ \ \ (a=1,3;b=\bar 2,\bar 4), \\
\lb{4e3}
& &\Big(G_{ab,0}K_{ab}\Big)_{\ \ \ v t}^{m p,}=G_{ab,0}K_{ab}
\Big[\frac{1}{2}\frac{1}{(N_c^{}+1)}
(\delta_{\ t}^{m}\delta_{\ v}^{p}+\delta_{\ v}^{m}\delta_{\ t}^{p})
-\frac{1}{2}\frac{1}{(N_c^{}-1)}
(\delta_{\ t}^{m}\delta_{\ v}^{p}-\delta_{\ v}^{m}\delta_{\ t}^{p})
\Big]\nonumber \\
& &\ \ \ \ \ \ \ \ \ \ \ \ (a=1,b=3,\ \mathrm{or},\ a=\bar 2,b=\bar 4),
\\
\lb{4e4}
& &K_{ab}=(-2\sigma D_{ab}) \ \ \ \ \ (a\neq b),
\eea
where the indices of $D$ correspond to the lines to which the gluon
propagator is connected.
\par
It can be checked that the iteration of the sum of the
kernels of two disjoint diagrams, for instance
$(G_{1\bar 2,0}K_{1\bar 2}+G_{3\bar 4,0}K_{3\bar 4})$, produces a
double counting with the interference term. This is cured at all
orders of perturbation theory by subtracting from the above kernel
the product of the two kernels
($G_{1\bar 2,0}K_{1\bar 2}G_{3\bar 4,0}K_{3\bar 4}$ for the example
above)
\cite{Huang:1974cd,Khvedelidze:1991qb,Kvinikhidze:2021kzu,
Yokojima:1993np,Bijtebier:2000xa,Heupel:2012ua,Eichmann:2020oqt,
Hoffer:2024fgm}.
\par
The total Green's function then satisfies the following integral
equation,
\bea \lb{4e5}
G_{nq,ki}^{mp,lj}&=&G_{nq,ki,0}^{mp,lj}
+\bigg\{\Big(\ (G_{1\bar 2,0}^{}4K_{1\bar 2}^{})
\mathbf{1}_{3\bar 4} 
  +\mathbf{1}_{1\bar 2}\,(G_{3\bar 4,0}^{}4K_{3\bar 4}^{})
  -(G_{1\bar 2,0}^{}4K_{1\bar 2}^{})\,(G_{3\bar 4,0}4K_{3\bar 4}^{})\
  \Big)_{nq,vt}^{mp,wu}\nonumber \\
& &\ \ \ \ \ +\Big(\ (G_{13,0}^{}4K_{13}^{})\mathbf{1}_{\bar 2\bar 4}
    +\mathbf{1}_{13}\,(G_{\bar 2\bar 4,0}^{}4K_{\bar 2\bar 4}^{})
    -(G_{13,0}^{}4K_{13}^{})\,(G_{\bar 2\bar 4,0}4K_{\bar 2\bar 4}^{})\
    \Big)_{nq,vt}^{mp,wu}\nonumber \\
& &\ \ \ \ \ +\Big(\ (G_{1\bar 4,0}^{}4K_{1\bar 4}^{})
    \mathbf{1}_{3\bar 2}+\mathbf{1}_{1\bar 4}\,
    (G_{3\bar 2,0}^{}4K_{3\bar 2}^{})
    -(G_{1\bar 4,0}^{}4K_{1\bar 4}^{})\,(G_{3\bar 2,0}4K_{3\bar 2}^{})\
    \Big)_{nq,vt}^{mp,wu}\bigg\}\,G_{uw,ki}^{tv,lj}.\nonumber\\
& &    
\eea
\par
We shall be interested in the color-singlet sectors of the incoming
and outgoing particles in the Green's function of Fig. \rf{f5}, which
are expected to be the only observable sectors on experimental grounds.
However, contrary to the case of ordinary mesons and baryons, color
singlets of multiquark states are not color irreducible,
in the sense that their color structure can be decomposed in general
along combinations of products of clusters with simpler color
structures, representing color singlets of quark-antiquark or
three-quark states
\cite{Jaffe:2008zz,Coleman:1985rnk,Nielsen:2009uh,Lucha:2019cdc}.
The decomposition of the Green's function (\rf{4e5}) along
components involving color-singlet tensors can be done in 
two ways. First, using the mesonic
clusters ($1\bar 2$)($3\bar 4$) and ($1\bar 4$)($3\bar 2$), or, 
second, using the diquark-antidiquark combinations
in their symmetric and antisymmetric representations. One
advantage of the first type of decomposition is its direct
relationship with the meson-meson scattering amplitudes, which are
other quantities of interest, while an advantage of the second type
of decomposition is its relationship with a possible existence of
compact tetraquark bound states. The two types of decomposition are
related to each other by linear transformations.
In the following, we shall consider the first type of
decomposition to establish the connection of the Green's function
with the meson-meson scattering amplitudes.
\par
Associating the incoming and outgoing states with the mesonic clusters,
one can distinguish four different channels in the transition process,
two of which will be called ``direct,'' the ingoing and outgoing
clusters being the same, and the two others ``recombination,''
the outgoing clusters having undergone a quark exchange.
The four different processes are,
\bea 
\lb{4e6}
& &(1\bar 2)+(3\bar 4)\ \longrightarrow (1\bar 2)+(3\bar 4),\ \ \ \
\mathrm{direct\ channel}\ 1\ \ \ (D1),\\
\lb{4e7}
& &(1\bar 2)+(3\bar 4)\ \longrightarrow (1\bar 4)+(3\bar 2),\ \ \ \
\mathrm{recombination\ channel}\ 1\ \ \ (R1),\\
\lb{4e8}
& &(1\bar 4)+(3\bar 2)\ \longrightarrow (1\bar 4)+(3\bar 2),\ \ \ \
\mathrm{direct\ channel}\ 2\ \ \ (D2),\\
\lb{4e9}
& &(1\bar 4)+(3\bar 2)\ \longrightarrow (1\bar 2)+(3\bar 4),\ \ \ \
\mathrm{recombination\ channel}\ 2\ \ \ (R2).
\eea
\par
Choosing a basis that projects, by contraction of color indices,
the Green's function on the above channels, one has the following
decomposition:
\bea \lb{4e10}
G_{nq,ki}^{mp,lj}=\frac{N_c^2}{(N_c^2-1)^2}\,&\bigg\{&\ \
(\delta_{\ n}^m\delta_{\ q}^p
-\frac{1}{N_c^{}}\delta_{\ q}^m\delta_{\ n}^p)
(\delta_{\ i}^j\delta_{\ k}^l
-\frac{1}{N_c^{}}\delta_{\ k}^j\delta_{\ i}^l)\ G_{D1}^{}
\nonumber \\
& &+(\delta_{\ q}^m\delta_{\ n}^p
-\frac{1}{N_c^{}}\delta_{\ n}^m\delta_{\ q}^p)
(\delta_{\ i}^j\delta_{\ k}^l
-\frac{1}{N_c^{}}\delta_{\ k}^j\delta_{\ i}^l)\ G_{R1}^{}
\nonumber \\
& &+(\delta_{\ q}^m\delta_{\ n}^p
-\frac{1}{N_c^{}}\delta_{\ n}^m\delta_{\ q}^p)
(\delta_{\ k}^j\delta_{\ i}^l
-\frac{1}{N_c^{}}\delta_{\ i}^j\delta_{\ k}^l)\ G_{D2}^{}
\nonumber \\
& &+(\delta_{\ n}^m\delta_{\ q}^p
-\frac{1}{N_c^{}}\delta_{\ q}^m\delta_{\ n}^p)
(\delta_{\ k}^j\delta_{\ i}^l
-\frac{1}{N_c^{}}\delta_{\ i}^j\delta_{\ k}^l)\ G_{R2}^{}\,\bigg\}
\ +\ \cdots,
\eea
where the dots stand for the existing higher representations than
the singlets. 
The normalization is such that
\be \lb{4e11}
\delta_{\ m}^n\delta_{\ p}^q\delta_{\ j}^i\delta_{\ l}^k\
G_{nq,ki}^{mp,lj}=N_c^2G_{D1}^{},\ \ \ \ \ \ 
\delta_{\ m}^q\delta_{\ p}^n\delta_{\ j}^i\delta_{\ l}^k\
G_{nq,ki}^{mp,lj}=N_c^2G_{R1}^{},
\ee
and similarly with the other two components, after appropriate
permutation of indices. The decomposition of the free Green's
function $G_{nq,ki,0}^{mp,lj}$ [Eq. (\rf{4e5})] is
\bea \lb{4e12}
G_{nq,ki,0}^{mp,lj}&=&\delta_{\ i}^m\delta_{\ n}^j
\delta_{\ k}^p\delta_{\ q}^l\,G_0\nonumber\\
&=&\frac{N_c^2}{(N_c^2-1)^2}\,\bigg\{\, 
(\delta_{\ n}^m\delta_{\ q}^p
-\frac{1}{N_c^{}}\delta_{\ q}^m\delta_{\ n}^p)
(\delta_{\ i}^j\delta_{\ k}^l
-\frac{1}{N_c^{}}\delta_{\ k}^j\delta_{\ i}^l)\,G_0^{}\nonumber\\
& &\ \ \ \ \ \ \ \ +(\delta_{\ q}^m\delta_{\ n}^p
-\frac{1}{N_c^{}}\delta_{\ n}^m\delta_{\ q}^p)
(\delta_{\ i}^j\delta_{\ k}^l
-\frac{1}{N_c^{}}\delta_{\ k}^j\delta_{\ i}^l)\,
\frac{1}{N_c^{}}G_0^{}\nonumber\\
& &\ \ \ \ \ \ \ \ +(\delta_{\ q}^m\delta_{\ n}^p
-\frac{1}{N_c^{}}\delta_{\ n}^m\delta_{\ q}^p)
(\delta_{\ k}^j\delta_{\ i}^l
-\frac{1}{N_c^{}}\delta_{\ i}^j\delta_{\ k}^l)\,G_0^{}\nonumber \\
& &\ \ \ \ \ \ \ \ +(\delta_{\ n}^m\delta_{\ q}^p
-\frac{1}{N_c^{}}\delta_{\ q}^m\delta_{\ n}^p)
(\delta_{\ k}^j\delta_{\ i}^l
-\frac{1}{N_c^{}}\delta_{\ i}^j\delta_{\ k}^l)\,
\frac{1}{N_c^{}}G_0^{}\,\bigg\}\ +\ \cdots.
\eea
\par
Taking into account the decompositions (\rf{4e10}) and (\rf{4e12})
and the kernel structures (\rf{4e2}) and (\rf{4e3}), the integral
equation (\rf{4e5}) is decomposed into two independent sets of
equations, each involving two coupled equations. The first set
concerns the components $G_{D1}^{}$ and $G_{R1}^{}$, the second set
$G_{D2}^{}$ and $G_{R2}^{}$. The latter set is obtained from the
former one by the interchange of the indices $\bar 2$ and
$\bar 4$. This is why we shall concentrate in the following 
on the first set only.
\par
Introducing the notation
\be \lb{4e13}
k_{ab}^{}\,\equiv 4\,G_{ab,0}^{}\,K_{ab}^{}\ \ \ \ \ \ (a\neq b),
\ee
[Eqs. (\rf{4e1}) and (\rf{4e4})], the coupled integral equations of
$G_{D1}^{}$ and $G_{R1}^{}$ take the following form:
\bea \lb{4e14}
G_{D1}^{}=G_0^{}
&+&(k_{1\bar 2}^{}+k_{3\bar 4}^{}
-k_{1\bar 2}^{}k_{3\bar 4}^{})
G_{D1}^{}\nonumber \\
&+&(k_{1\bar 4}^{}+k_{3\bar 2}^{})\,
\frac{N_c^{}}{(N_c^2-1)}(-\frac{1}{N_c^{}}G_{D1}^{}+G_{R1}^{})
\nonumber \\
&-&k_{1\bar 4}^{}k_{3\bar 2}^{}\,
\Big[\,\frac{1}{N_c^{}}G_{R1}^{}+\frac{1}{(N_c^2-1)^2}
(G_{D1}^{}-\frac{1}{N_c^{}}G_{R1}^{})\,\Big]\nonumber \\
&+&(k_{13}^{}+k_{\bar 2\bar 4}^{})\,
\frac{N_c^{}}{(N_c^2-1)}(-\frac{1}{N_c^{}}G_{D1}^{}+G_{R1}^{})
\nonumber \\
&-&k_{13}^{}k_{\bar 2\bar 4}^{}\,
\frac{N_c^2}{(N_c^2-1)^2}
\Big[\,(1+\frac{1}{N_c^2})G_{D1}^{}-\frac{2}{N_c^{}}G_{R1}^{}\,\Big].  
\eea
\bea \lb{4e15}
G_{R1}^{}=\frac{1}{N_c^{}}G_0^{}
&+&(k_{1\bar 4}^{}+k_{3\bar 2}^{}
-k_{1\bar 4}^{}k_{3\bar 2}^{})
G_{R1}^{}\nonumber \\
&+&(k_{1\bar 2}^{}+k_{3\bar 4}^{})\,
\frac{N_c^{}}{(N_c^2-1)}(G_{D1}^{}-\frac{1}{N_c^{}}G_{R1}^{})
\nonumber\\
&-&k_{1\bar 2}^{}k_{3\bar 4}^{}\,
\Big[\,\frac{1}{N_c^{}}G_{D1}^{}-\frac{1}{(N_c^2-1)^2}
(\frac{1}{N_c^{}}G_{D1}^{}-G_{R1}^{})\,\Big]\nonumber \\
&+&(k_{13}^{}+k_{\bar 2\bar 4}^{})\,
\frac{N_c^{}}{(N_c^2-1)}(G_{D1}^{}-\frac{1}{N_c^{}}G_{R1}^{})
\nonumber\\
&-&k_{13}^{}k_{\bar 2\bar 4}^{}\,
\frac{N_c^2}{(N_c^2-1)^2}
\Big[\,-\frac{2}{N_c^{}}\,G_{D1}^{}+(1+\frac{1}{N_c^2})G_{R1}^{}\,
\Big].  
\eea
\par
Let us notice that one-gluon exchanges between two color-singlet
states are generally forbidden by color conservation. Here, this
process concerns the Green's function $G_{D1}^{}$, which describes
the interaction between the color-singlet states $(1\bar 2)$ and
$(3\bar 4)$ [Eq. (4.6)]. It can be checked that the
one-gluon-exchange terms between the two states that appear through
the quantities
$(k_{1\bar 4}^{}+k_{3\bar 2}^{})$ and
$(k_{13}^{}+k_{\bar 2\bar 4}^{})$,
do disappear in lowest order of perturbation theory, when one replaces
in the accompanying terms $G_{D1}^{}$ and $G_{R1}^{}$ by their
lowest-order expressions $G_0^{}$ and $G_0^{}/N_c^{}$, respectively.
\par

\subsection{Scattering amplitudes} \lb{s42}

To obtain the integral equations of the scattering amplitudes of the
meson clusters, one first has to restore the $\gamma$ matrix
structure of the Green's function equations, as in the two-body
case [before Eq. (\rf{2e43})]. Because of the particular structure
of the kernels $K_{ab}^{}$ in $\gamma$ matrices (tensor products
$\gamma_-^{}\times\gamma_-^{}$) and the dynamical role of the
components $G_{-,-,-,-}^{}$ of the Green's functions, the
scattering amplitudes are tensor products of four $\gamma_-^{}$
matrices. After contraction of
the $\gamma$ matrices among themselves, one comes back to the basis
that was used for the Green's functions [cf. Eqs. (\rf{4e1})].
\par
The off-mass shell scattering amplitudes of the processes
(\rf{4e6})-(\rf{4e9}) are obtained by subtracting the disconnected
parts due to the corresponding clusters in the direct channels and
then factorizing out the related two-body Green's functions. We
adopt for the channels $D1$ and $R1$, in the projected space of the
$\gamma$ matrices, the definitions,
\bea
\lb{4e16}
& &G_{D1}^{}=G_{1\bar 2}^{}G_{3\bar 4}^{}+
G_{1\bar 2}^{}G_{3\bar 4}^{}\,\mathcal{T}_{D1}^{}\,
G_{1\bar 2}^{}G_{3\bar 4}^{},\\
\lb{4e17}
& &G_{R1}^{}=G_{1\bar 4}^{}G_{3\bar 2}^{}\,\mathcal{T}_{R1}^{}\,
G_{1\bar 2}^{}G_{3\bar 4}^{}.
\eea
In replacing these expressions in Eqs. (\rf{4e14}) and (\rf{4e15}),
one uses in a few $N_c^{}$-leading terms the equations satisfied by
$G_{1\bar 2}^{}G_{3\bar 4}^{}$ and $G_{1\bar 4}^{}G_{3\bar 2}^{}$,
namely [Eq. (\rf{2e30})]
\be \lb{4e18}
(1-k_{1\bar 2}^{})(1-k_{3\bar 4}^{})G_{1\bar 2}^{}G_{3\bar 4}^{}
=G_0^{},\ \ \ \ \ \ 
(1-k_{1\bar 4}^{})(1-k_{3\bar 2}^{})G_{1\bar 4}^{}G_{3\bar 2}^{}
=G_0^{}.
\ee
\par
One obtains,
\bea \lb{4e19}
G_0^{}\mathcal{T}_{D1}^{}G_0^{}&=&-(k_{1\bar{4}}^{}+k_{3\bar{2}}
+k_{13}^{}+k_{\bar{2}\bar{4}})
\frac{1}{(N_c^2-1)}\Big[1+G_{1\bar{2}}^{}G_{3\bar{4}}
  \mathcal{T}_{D1}^{}-N_c^{}G_{1\bar{4}}^{}G_{3\bar{2}}
  \mathcal{T}_{R1}^{}
\Big]\,G_0^{}\nonumber \\
& &-k_{1\bar{4}}^{}k_{3\bar{2}}\Big[\frac{1}{N_c^{}}
G_{1\bar{4}}^{}G_{3\bar{2}}\mathcal{T}_{R1}^{}+\frac{1}{(N_c^2-1)^2}
\Big(1+G_{1\bar{2}}^{}G_{3\bar{4}}\mathcal{T}_{D1}^{}
-\frac{1}{N_c^{}}G_{1\bar{4}}^{}G_{3\bar{2}}\mathcal{T}_{R1}^{}\Big)
\Big]\,G_0^{}\nonumber \\
& &-k_{13}^{}k_{\bar{2}\bar{4}}\frac{1}{(N_c^2-1)^2}\Big[
(N_c^2+1)
+(N_c^2+1)G_{1\bar{2}}^{}G_{3\bar{4}}\mathcal{T}_{D1}^{}
-2N_c^{}G_{1\bar{4}}^{}G_{3\bar{2}}\mathcal{T}_{R1}^{}\Big]\,G_0^{}.
\nonumber \\
& &
\eea
\bea \lb{4e20}
G_0^{}\mathcal{T}_{R1}^{}G_0^{}&=&\frac{G_0^{}}{N_c^{}}
+\frac{1}{N_c^{}}(k_{1\bar{2}}^{}+k_{3\bar{4}}
-k_{1\bar{2}}^{}k_{3\bar{4}})
G_{1\bar{2}}^{}G_{3\bar{4}}\mathcal{T}_{D1}^{}\,G_0^{}\nonumber \\
& &+\frac{1}{(N_c^2-1)}\Big(k_{1\bar{2}}^{}+k_{3\bar{4}}
+\frac{1}{(N_c^2-1)}k_{1\bar{2}}^{}k_{3\bar{4}}\Big)
\Big(\frac{1}{N_c^{}}+\frac{1}{N_c^{}}G_{1\bar{2}}^{}G_{3\bar{4}}
\mathcal{T}_{D1}^{}-G_{1\bar{4}}^{}G_{3\bar{2}}\mathcal{T}_{R1}^{}
\Big)\,G_0^{}\nonumber \\
& &+(k_{13}^{}+k_{\bar{2}\bar{4}})\Big[\frac{N_c^{}}{(N_c^2-1)}
(1+G_{1\bar{2}}^{}G_{3\bar{4}}\mathcal{T}_{D1}^{})
-\frac{1}{(N_c^2-1)}G_{1\bar{4}}^{}G_{3\bar{2}}\mathcal{T}_{R1}^{}  
\Big]\,G_0^{}\nonumber \\
& &-k_{13}^{}k_{\bar{2}\bar{4}}\frac{1}{(N_c^2-1)^2}\Big[
-2N_c^{}-2N_c^{}G_{1\bar{2}}^{}G_{3\bar{4}}
\mathcal{T}_{D1}^{}
+(N_c^2+1)G_{1\bar{4}}^{}G_{3\bar{2}}\mathcal{T}_{R1}^{}\Big]\,G_0^{}.
\nonumber \\
& &
\eea
The two-body Green's functions $G_{ab}^{}$ satisfy in turn Eqs.
(\rf{2e49}),
\be \lb{4e21}  
G_{ab}^{}=G_{ab,0}^{}+G_{ab,0}^{}\,4\mathcal{T}_{ab}\,G_{ab,0}^{}
\ \ \ \ \ \ \ (a\neq b).
\ee
[The factor $1/N_c^{}$ has been removed from Eq. (\rf{2e49}), being
now included in the normalization conditions (\rf{4e11}) and
(\rf{4e12}) of the four-body Green's functions; $\mathcal{T}_{ab}$
has expression (\rf{2e45}) without the factors $1/N_c^{}$.]
\par
The on-mass shell scattering amplitudes are obtained by projecting
Eqs. (\rf{4e19}) and (\rf{4e20}) on the two-body wave functions of
the type of (\rf{2e46}), with factors $2$ [Eq. (\rf{2e48}) without
the factor $1/\sqrt{N_c^{}}$], and imposing on the corresponding
total momenta the meson mass-shell constraints. Thus,
\be \lb{4e22}
\Big(\mathcal{T}_{D1}^{}\Big)_{nm,n'm'}=2^4\,
\widetilde{\phi}_{1\bar{2},n}^*\widetilde{\phi}_{3\bar{4},m}^*
\Big(G_0^{}\mathcal{T}_{D1}^{}G_0^{}\Big)
\widetilde{\phi}_{1\bar{2},n'}^{}\widetilde{\phi}_{3\bar{4},m'}^{},
\ee
\be \lb{4e23}
\Big(\mathcal{T}_{R1}^{}\Big)_{pq,n'm'}=2^4\,
\widetilde{\phi}_{1\bar{4},p}^*\widetilde{\phi}_{3\bar{2},q}^*
\Big(G_0^{}\mathcal{T}_{R1}^{}G_0^{}\Big)
\widetilde{\phi}_{1\bar{2},n'}^{}\widetilde{\phi}_{3\bar{4},m'}^{},
\ee
where $n,m,p,q,n',m'$ are the quantum numbers of the meson states.
\par
Equations (\rf{4e19}) and (\rf{4e20}) display the explicit dependence
on $N_c^{}$ of the scattering amplitudes. The recombination scattering
amplitude $\mathcal{T}_{R1}^{}$ is of order $N_c^{-1}$, while
$\mathcal{T}_{D1}^{}$ is of order $N_c^{-2}$, with subleading orders
decreasing by orders of $N_c^{-2}$ in both cases. The leading-order
term of $\mathcal{T}_{D1}^{}$ is generated by the iteration of the
leading-order term of $\mathcal{T}_{R1}^{}$, a feature that is also
valid in four dimensions, noticed from the analysis  of the structure
of the corresponding Feynman diagrams
\cite{Lucha:2021mwx,Lucha:2017gqq}.
\par
At leading orders one has
\bea
\lb{4e24}
G_0^{}\mathcal{T}_{R1}^{}G_0^{}&=&\frac{1}{N_c^{}}
(1+k_{13}^{}+k_{\bar{2}\bar{4}})G_0^{}+O(N_c^{-3}),\\
\lb{4e25}
G_0^{}\mathcal{T}_{D1}^{}G_0^{}&=&-\frac{1}{N_c^2}
(k_{1\bar{4}}^{}+k_{3\bar{2}}+k_{13}^{}+k_{\bar{2}\bar{4}})
\Big[1-N_c^{}G_{1\bar{4}}^{}G_{3\bar{2}}
\mathcal{T}_{R1}^{}\Big]\,G_0^{}\nonumber \\
& &-k_{1\bar{4}}^{}k_{3\bar{2}}\,\frac{1}{N_c^{}}
G_{1\bar{4}}^{}G_{3\bar{2}}\mathcal{T}_{R1}^{}G_0^{}
-\frac{1}{N_c^2}k_{13}^{}k_{\bar{2}\bar{4}}^{}G_0^{}
+O(N_c^{-4}).  
\eea
\par
We notice that, as in the case of $G_{D1}^{}$ [Eq. (\rf{4e14})],
the terms in the brackets in Eq. (\rf{4e25}) are themselves
proportional to interaction terms, after $\mathcal{T}_{R1}^{}$
has been replaced by its lowest-order expression (\rf{4e24})
and decompositions (\rf{4e21}) are used, and therefore
$\mathcal{T}_{D1}^{}$ does not contain one-gluon exchange terms.
\par

\section{Finiteness of the scattering amplitudes to order
\boldmath{$1/N_{\lowercase{c}}^2$}} \lb{s5}

One of the important tests of the present theory is the check that
physical observables are finite quantities, independent of the
infrared cutoff introduced in intermediate calculations. This
was verified in the case of the quark-antiquark system for
color-singlet states (Sec. \rf{s2}). The meson-meson scattering
amplitudes on the mass-shell being also observable quantities, one
must also check their independence of the cutoff parameter. The
$N_c^{}$-leading-order expressions of the scattering amplitudes,
given in Eqs. (\rf{4e24}) and (\rf{4e25}), allow us to proceed to
this check and to extract from them the physical content they
convey.
\par
Another check, which goes in parallel to the previous one, is the
verification that the meson-meson scattering amplitudes, even if
they are finite, are free of long-range van der Waals type forces.
This is not evident from Eqs. (\rf{4e24}) and (\rf{4e25}), since
they exhibit many gluon propagators combined with various other
quantities.
\par
We shall mainly present in this section the qualitative aspects
of the calculations with their final results, leaving the technical
details to Appendix \rf{sa2}.
It is understood that expressions (\rf{4e24}) and (\rf{4e25})
are projected, as in Eqs. (\rf{4e22}) and (\rf{4e23}), on the
meson wave functions; however, for the simplicity of notation,
we shall often omit them in the formulas.
\par
We start with the recombination scattering amplitude (\rf{4e24}).
It is composed of three contributions, which are graphically
represented in Fig. \rf{f6}.  
\bfg 
\vspace*{1 cm}
\bc
\hspace{-2 cm}
\includegraphics[scale=0.7]{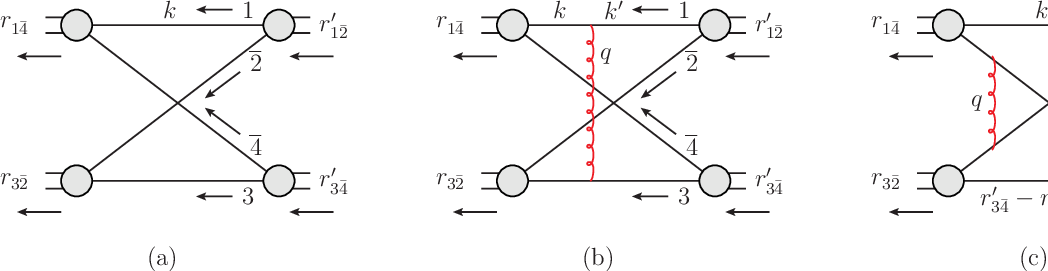}
\caption{Three diagrams contributing to the recombination
scattering amplitude $R1$ [Eq. (\rf{4e7})] at order $1/N_C^{}$.}
\lb{f6} 
\ec
\efg
\par
In this figure, the circles represent the external meson wave
functions, such as $\widetilde{\phi}_{1\bar{2}}^{}$ and
$\widetilde{\phi}_{3\bar{4}}^{}$ for the incoming ones, and
$\widetilde{\phi}_{1\bar{4}}^*$ and $\widetilde{\phi}_{3\bar{2}}^*$
for the outgoing ones; in the present discussion, the specific bound
state quantum numbers of the wave functions are irrelevant and are
omitted. The arrows indicate the momentum routings of the quarks and
antiquarks; the flavors of the quarks and antiquarks are also
indicated in the internal lines. Solid lines represent quark and
antiquark propagators and the curly lines the gluon propagator.
The loop momenta are indicated on some quark lines and the whole
momentum distributions can then easily be reconstituted from the
knowledge of the external momenta, which are the only physical
momenta in the scattering processes. Meson wave functions will be
specified by the total momentum they carry and by their internal
quark momentum. The wave functions $\widetilde{\phi}$ have a pair
of quark and antiquark outgoing toward the interior of the diagrams,
while the $\widetilde{\phi}^*$s have an incoming pair from the
interior of the diagrams.
Thus the wave function $\widetilde{\phi}_{1\bar{2}}^{}$ in diagram
(a) has total momentum $r_{1\bar 2}'$ and quark momentum $k$;
the wave function $\widetilde{\phi}_{1\bar{4}}^{*}$ has total
momentum $r_{1\bar 4}'$ and quark momentum $k$.
In diagrams (b) and (c), the momentum carried by the gluon is
$q=k'-k$.
\par
One important remark concerns the spectral conditions imposed on
the wave functions. It is assumed that the external mesons have
physical momenta, which means that the components $r_{a\bar b,-}$
of their total momentum $r_{a\bar b}$ are positive. On the other
hand, it is seen from Eq. (\rf{2e47}) that, for the divergent part of
the wave function, each quark and antiquark should also separately
satisfy such a condition. When the latter condition is not satisfied
at the quark and antiquark levels, it is the subleading part in
$\Lambda$ of the wave function that contributes. However, for the
class of quark flavors that we are considering in the present work
(four different flavors), the same phenomenon also occurs at a second
wave function connected to the previous one with the concerned quark
or antiquark lines. For instance, if, in diagram (a) of Fig. \rf{f6},
$k_-<0$, then this feature will concern the two wave functions
$\widetilde{\phi}_{1\bar{2}}^{}$ and $\widetilde{\phi}_{1\bar{4}}^*$,
and, therefore, the reduction of the power in $\Lambda$ will be of
two degrees. As the detailed calculations show, such a reduction
eliminates the corresponding contributions from the physical quantity
when the limit $\Lambda\rightarrow\infty$ is taken. The same
phenomenon also occurs in the presence of gluon propagators. For
instance, in diagram (b) of Fig. \rf{f6}, one may have $k_-'<0$; the
leading divergence in the diagram will come from the singularity
of the gluon propagator, which implies the limit $k_-'=k_-$, and thus
the contribution of the quark propagator reaching the meson
$(1\bar 4)$ will also reduce the power of $\Lambda$ by one degree;
one again falls in the case considered above.
Therefore, one may calculate the leading and subleading powers of
$\Lambda$ by retaining only that part of each wave function
[Eq. (\rf{2e47})] that satisfies the above spectral conditions.
\par
In diagrams having gluon propagators, the leading divergences will
receive contributions from the singularities of the latter. 
As in Eqs. (\rf{2e34}) and (\rf{2e35}), the finite parts of the
gluon propagators will contribute to the subleading terms in
$\Lambda$ of the diagrams.
\par
We can now evaluate the contributions of the diagrams of Fig.
\rf{f6}. Diagram (a) contains four quark propagators, each
carrying a damping factor $\Lambda^{-1}$ [Eq. (\rf{2e18})]. The
global damping factor is therefore $\Lambda^{-4}$. However, these
propagators are submitted to the integration with respect to the
component $k_+^{}$ of the loop momentum, which transforms the
damping factor into $\Lambda^{-3}$. Taking into account the
$\Lambda^4$ diverging factor coming from the external wave
functions, one finds a final diverging factor $\Lambda^1$, plus
additional finite contributions. The diagram is therefore divergent
and could not account alone for the description of a physical
process. One therefore has to associate with it the two other
diagrams, (b) and (c), of Fig. \rf{f6}.
\par
Diagram (b) has six quark propagators and an associated damping
factor $\Lambda^{-6}$. There are two loop integrations with respect
to $k_+^{}$ and $k_+'$, which transform $\Lambda^{-6}$ into
$\Lambda^{-4}$. The singularity of the gluon propagator increases
$\Lambda^{-4}$ up to $\Lambda^{-3}$. With the contribution of the
external wave functions, this is transformed into $\Lambda^{1}$.
A similar result is also obtained with diagram (c).
\par
It turns out that the sum of the divergences coming from the three
diagrams vanishes and the global result is a finite quantity,
coming from the finite parts of the quark and gluon propagators
and from the instantaneous part, $\varphi$, of the wave functions
[Eqs. (\rf{2e32}) and (\rf{2e47})]. The details of the above
calculations can be found in Appendix \rf{sa2}. We simply describe
here the main qualitative features of the resulting finite expression
of the scattering amplitude [cf. the discussion after
Eq. (\rf{a2e10}), leading to Eq. (\rf{a2e11})].
\par
The latter can be expressed as a double integral, in the $-$
components of the loop momenta, within finite bounds, involving the
four external wave functions and the finite part of the gluon
propagator. The expression does not display any singularity in the
momentum transfer of the scattering process, which would be the
signal of the existence of long-range van der Waals type forces.
Rather, it displays a smooth dependence on the momenta, dominated
by a constant term, of the scale of the string tension $\sigma$.
The domains of the spectral conditions of the four meson wave
functions do not generally coincide in the integrals and lead to
complicated momentum-dependent expressions, which could be determined
only by detailed numerical calculations. It is the nonsingular
property of these expressions that allows us to approximate them by
a local expression [Eq. (\rf{a2e11})], represented by its scalar,
momentum-independent, part.
In this respect, it can be viewed as a generalized effective contact
term. This is represented in Fig. \rf{f7}.
\bfg 
\bc
\includegraphics[scale=0.85]{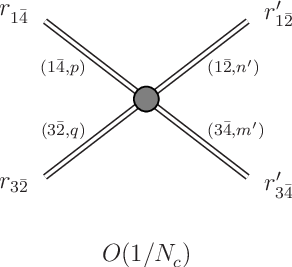}
\caption{Effective contact interaction representing the
recombination scattering amplitude $R1$ at order $1/N_c^{}$.
The flavor and excitation quantum numbers of the mesons are
also indicated.} 
\lb{f7} 
\ec
\efg
\par
We next consider the direct process represented in Eq. (\rf{4e25}).
Here, the right-hand side contains the two-particle propagators
$G_{1\bar{4}}^{}$ and $G_{3\bar{2}}$, whose structure in terms of the
two-body scattering amplitudes is displayed in Eq. (\rf{4e21}).
The latter, in turn, have the structure represented in Eq.
(\rf{2e45}), composed of the contributions of one-gluon exchange
and meson-bound-state poles. We separate in a more explicit form
the latter contributions by adopting the following notation:
\be \lb{5e1}
G_{ab,0}^{}4\mathcal{T}_{ab}^{}\equiv k_{ab}^{}+
G_{ab,0}^{}4\widetilde{\mathcal{T}}_{ab}^{},
\ee
where $k_{ab}^{}$ is defined in Eqs. (\rf{4e4}) and (\rf{4e13}) and
$\widetilde{\mathcal{T}}_{ab}^{}$ contains the contributions
of the meson poles. Equation (\rf{4e25}) can then be expressed in
the following form:
\be \lb{5e2}
N_c^2G_0^{}\mathcal{T}_{D1}^{}G_0^{}=\
\Big\{(a)+(b)+(c)+(d)+(e)\Big\},
\ee
where the symbols in the right-hand side have the expressions
\bea
\lb{5e3}
(a)&=&(1+k_{13}^{}+k_{\bar 2\bar 4}^{})\,
k_{1\bar 4}^{}k_{3\bar 2}^{}\,
(1+k_{13}^{}+k_{\bar 2\bar 4}^{})\,G_0^{},\\
\lb{5e4}
(b)&=&(1+k_{13}^{}+k_{\bar 2\bar 4}^{})\,(k_{1\bar 4}^{}
G_{3\bar 2,0}^{}
4\widetilde{\mathcal{T}}_{3\bar 2}^{}+k_{3\bar 2}^{}G_{1\bar 4,0}^{}
4\widetilde{\mathcal{T}}_{1\bar 4}^{})\,
(1+k_{13}^{}+k_{\bar 2\bar 4}^{})\,G_0^{},\\
\lb{5e5}
(c)&=&(1+k_{13}^{}+k_{\bar 2\bar 4}^{})\,G_{3\bar 2,0}^{}
4\widetilde{\mathcal{T}}_{3\bar 2}^{}\,G_{1\bar 4,0}^{}
4\widetilde{\mathcal{T}}_{1\bar 4}^{}\,
(1+k_{13}^{}+k_{\bar 2\bar 4}^{})\,G_0^{},\\
\lb{5e6}
(d)&=&(1+k_{13}^{}+k_{\bar 2\bar 4}^{})\,(G_{3\bar 2,0}^{}
4\widetilde{\mathcal{T}}_{3\bar 2}^{}+G_{1\bar 4,0}^{}
4\widetilde{\mathcal{T}}_{1\bar 4}^{})\,
(1+k_{13}^{}+k_{\bar 2\bar 4}^{})\,G_0^{},\\
\lb{5e7}
(e)&=&(1+k_{13}^{}+k_{\bar 2\bar 4}^{})\,
(k_{1\bar 4}^{}+k_{3\bar 2}^{})\,
(1+k_{13}^{}+k_{\bar 2\bar 4}^{})\,G_0^{}\nonumber \\
& &\ \ \ \ \ +\Big[\,(k_{13}^2+k_{\bar 2\bar 4}^2
+k_{13}^{}k_{\bar 2\bar 4}^{})-(k_{1\bar 4}^{}+k_{3\bar 2}^{})\,\Big]
\,G_0^{}.
\eea
\par
We first consider expression (c), Eq. (\rf{5e5}). It contains nine
different contributions. The part not containing gluon propagators
is graphically represented in Fig. \rf{f8}.
\par
\bfg 
\bc
\includegraphics[scale=0.7]{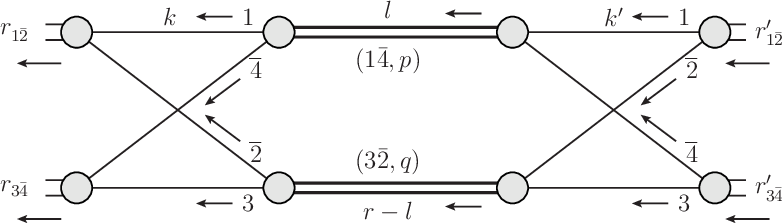}
\caption{Diagrammatic representation of the part of expression (c),
Eq. (\rf{5e5}), not containing gluon propagators. Double lines
represent meson propagators. $r$ is the total momentum.}
\lb{f8} 
\ec
\efg
\par
This diagram is made of three parts: a central part, containing
two meson propagators with quark-antiquark content $(1\bar 4)$ and
$(3\bar 2)$, respectively, and summed over the infinite number of
mesons appearing in the quark-antiquark spectrum [Eq. (\rf{2e45})],
and two side parts, corresponding to the recombination diagram (a) of
Fig. \rf{f6}. The other parts of Eq. (\rf{5e5}) correspond to the
inclusion of the gluon propagator in the side parts, as in diagrams
(b) and (c) of Fig. \rf{f6}. Therefore, as is also evident from
Eq. (\rf{4e24}), expression (\rf{5e5}) represents the inclusion of
the two meson propagators between the two recombination amplitudes
$R2$ and $R1$ [Eqs. (\rf{4e9}) and (\rf{4e7})], each at order
$1/N_c^{}$. It is not necessary
in this case to calculate separately the contribution of each
diagram. For a fixed value of the momentum $l$, which
appears in the meson propagators, the integration of the momenta of
the left and right sets of recombination diagrams can be done
independently, provided the momentum $l$ is considered as part of
the external momenta. As we have seen previously, each set of the
three recombination diagrams leads to a finite result, represented
by the generalized contact-type interaction of Fig. \rf{f7}.
At this stage, one should not yet take the limit
$\Lambda\rightarrow\infty$, awaiting the integration with respect
to the component $+$ of the momentum $l$. The latter operation
yields the once-integrated (with respect to $l_+^{}$) finite
expression of the two-meson loop amplitude, taken between the
two effective contact terms of the recombination amplitudes.
The result is therefore equivalent to the generalized form of
the unitarity diagram, including the infinite number of intermediate
mesons, and generated by the recombination process at order $1/N_c$,
the whole contribution being of order $1/N_c^2$. This is represented
in Fig. \rf{f9}.
\par
\bfg 
\bc
\includegraphics[scale=0.85]{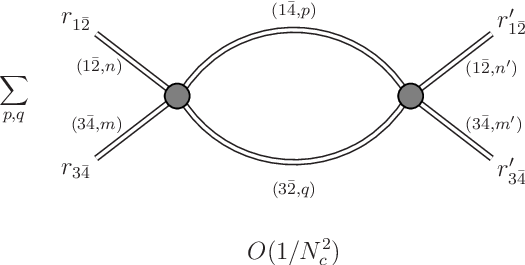}
\caption{Effective unitarity diagrams arising from expression
(\rf{5e5}) in the direct scattering amplitude $D1$ at order
$1/N_c^2$. The flavor and excitation quantum numbers of the mesons
are also indicated.}
\lb{f9} 
\ec
\efg
\par
We next consider expression (a), Eq. (\rf{5e3}). It is composed
of nine diagrams, with quark and gluon propagators, not involving
mesons. A typical diagram of this category is represented in
Fig. \rf{f10}. Here we encounter diagrams that may contain
several gluon lines.
\bfg 
\bc
\includegraphics[scale=0.8]{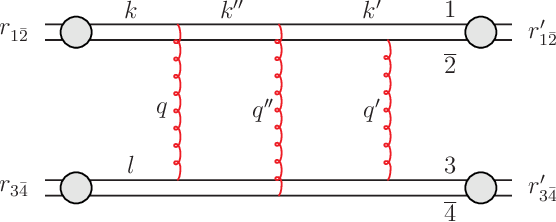}
\caption{Typical diagram contributing to Eq. (\rf{5e3}), with 
three gluon exchanges, corresponding to $k_{13}^{}$, $k_{1\bar 4}^{}$,
and $k_{3\bar 2}^{}$. The momentum routings of the quarks and the
antiquarks are from right to left.}  
\lb{f10} 
\ec
\efg
\par
One of the main properties in the diagrams
of the direct processes $D1$ and $D2$, except for the category (c)
that we have met earlier, is that the sum of the momenta of the
exchanged gluons is equal to the momentum transfer
$(r_{1\bar 2,-}'-r_{1\bar 2,-}^{})$; as a consequence, when the
leading singularities of the gluon propagators are considered for
the evaluation of the leading effects in $\Lambda$, one of the gluon
propagators factorizes out of the integrals with the expression
$-2i\sigma/(r_{1\bar 2,-}'-r_{1\bar 2,-}^{})^2$.
Therefore, the gluon propagators
provide singularities with one degree less in $\Lambda$ than
superficially expected. The whole contribution of the diagram
becomes in that case finite. However, the counterpart of this
result is that one faces now an expression where an explicit
gluon propagator is present, as if the interaction between the
mesons was a confining one. This, of course, would be unacceptable,
since the mutual meson interactions not only are not confining, but
also are expected to be free of long-range van der Waals type forces.
It is therefore mandatory, as for the infrared diverging quantities,
to verify that the final sum of such quantities does indeed vanish.
The details of  the calculation of the contributions in the case of
several gluon propagators is presented in Appendixes \rf{sa1} and
\rf{sa2}. 
\par
Coming back to the example of Fig. \rf{f10} and to the counting
rules of the damping and diverging factors, we have there ten
quark propagators, yielding a damping factor $\Lambda^{-10}$.
The four loop momenta increase, by integration with respect to
the $+$ components, this factor up to $\Lambda^{-6}$. Among the
three gluon propagators, only two of them provide diverging
singularities, transforming the latter factor into $\Lambda^{-4}$,
which itself is balanced by the diverging factor $\Lambda^{4}$
coming from the external wave functions. The final quantity is
therefore finite with the explicit presence of a gluon propagator
carrying the momentum transfer of the process.
It is verified that the sum of leading singularities of the nine
contributions of expression (a), Eq. (\rf{5e3}), vanishes in the
limit $\Lambda\rightarrow\infty$. One is left with finite nonsingular
contributions, which induce a generalized effective contact-type
interaction at the mesonic level.
\par
The next category in Eq. (\rf{5e2}) is expression (b), Eq. (\rf{5e4}),
containing 18 contributions. A typical diagram of it is represented in
Fig. \rf{f11}. Here, we have the presence of only one type of
meson propagators.
\bfg 
\bc
\includegraphics[scale=0.8]{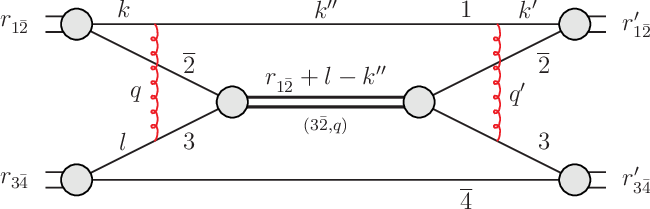}
\caption{Typical diagram contributing to Eq. (\rf{5e4}), with 
two gluon exchanges, each corresponding to $k_{13}^{}$.
Double lines represent meson propagators, whose flavor and excitation
quantum numbers are also indicated. The momentum routings of the
quarks and the antiquarks are from right to left.}   
\lb{f11} 
\ec
\efg
\par
In Fig. \rf{f11}, we have ten quark propagators, yielding the
damping factor $\Lambda^{-10}$. We have four loop momenta, which,
however, produce only a diverging factor $\Lambda^{3}$, because of
the presence of the meson propagator, which recuperates a damping
factor $\Lambda^{-1}$ (or prevents one damping factor from
disappearing from a quark propagator). Among the two gluon
propagators, only one of them produces a diverging singularity,
transforming the damping factor into $\Lambda^{-6}$, which is
balanced by the diverging factor $\Lambda^{6}$ coming from the six
wave functions. One remains with the explicit presence of a gluon
propagator, carrying the momentum transfer of the process. It is
verified that the sum of leading singularties of the contributions
of the nine diagrams, grouped around the present meson propagator
[$(3\bar 2,q)$], and that of the other nine diagrams, grouped around
the meson propagator $(1\bar 4,p)$, vanish separately, leaving
finite regular quantities, which induce generalized contact-type
interactions at the mesonic level.
\par
The expressions (d) and (e) of Eq. (\rf{5e2}), given in Eqs.
(\rf{5e6}) and (\rf{5e7}), have structures that are very similar
to those of expressions (b) and (a), respectively, from which they
are distinguished only by the difference in the connections of the
gluon propagators. The same counting rules are applied and again a
factorized gluon propagator is found. One again finds for the
sum of contributions in expressions (d) and (e) zero for the
singular parts, with finite regular remainders leading to contact
terms. 
\par
The resulting generalized contact term of the direct channel is
represented in Fig. \rf{f12}.
\bfg 
\bc
\includegraphics[scale=0.8]{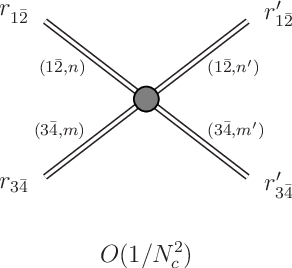}
\caption{Effective contact interaction resulting in the
direct scattering amplitude $D1$ at order $1/N_c^{2}$.
The flavor and excitation quantum numbers of the mesons are
also indicated.} 
\lb{f12} 
\ec
\efg
\par
In summary, to order $1/N_c^2$, the meson-meson scattering amplitudes
reduce, after the infrared cutoff is taken to zero, to three effective
finite quantities: (1) and (2), generalized contact terms in the
recombination and direct channels (Figs. \rf{f7} and \rf{f12}),
of order $1/N_c^{}$ and $1/N_c^{2}$, respectively; (3), a sum of
unitarity diagrams in the direct channels (Fig. \rf{f9}), of order
$1/N_c^2$, generated by the recombination-channel contact terms
and the meson propagators.
\par

\section{The scattering amplitudes in unitarized form} \lb{s6}

The check of the finiteness of the scattering amplitudes at
higher orders in $1/N_c^{}$ is not a straightforward task,
because of the continuous increasing of the number of the
diagrams and the appearance of new categories of topologies.
The latter, however, are expected to have weaker impact than
those met above at large $N_c^{}$. It is reasonable at this stage
to stick to the approximation that consists in selecting from
higher-order terms those which contribute to the unitarization of
the scattering amplitudes. We already found that the theory has
the tendency to reach that structure, by providing in the direct
channels the unitarity diagrams generated by the recombination
channel contact terms.
\par
We consider that procedure by adopting two approximations. First,
we consider for the external mesons the ground state mesons and we
neglect in the unitarity diagrams the contributions of higher excited
mesons. This is justified by the facts that the latter contribute
with larger mass damping factors and have smaller contributions to
the contact terms, due to the overlapping integrals between wave
functions of the ground states and their radial excitations.
Second, the contact terms are considered in their scalar
approximation, which is expected to provide the dominant
contribution, and are treated as constants. With these
approximations, the unitarization procedure amounts to dealing with
four coupled equations, which can be treated in matrix form.
\par
Before proceeding further, we shall slightly change the convention
of the scattering amplitude with respect to the one adopted earlier,
by explicitly factorizing the coefficient $i$ from it, as well as in
the kernels of the related integral equations,
\be \lb{6e1}
\mathcal{T}=i\mathcal{T}'\equiv i\mathcal{T}.
\ee
We shall continue using the same notation $\mathcal{T}$ for the
newly defined one.
\par
Designating by $K_{R1}^{}$ and $K_{R2}^{}$ the real contact terms
of the recombination channels (unitarity requiring the equality
$K_{R1}^{}=K_{R2}^{}$),
by $K_{D1}^{}$ and $K_{D2}$ those of the direct channels, and
by $J_{1\bar 2,3\bar 4}^{}\equiv J_{D1}^{}$ and
$J_{1\bar 4,3\bar 2}^{}\equiv J_{D2}^{}$ the
two-meson loop functions, the scattering amplitudes and the kernels
of the integral equations they satisfy can be represented in the
following matrix forms:
\be \lb{6e2}
\mathcal{T}=\left( \ba{cc}
\mathcal{T}_{D1}^{} & \mathcal{T}_{R2}^{} \\
\mathcal{T}_{R1}^{} & \mathcal{T}_{D2}^{}
\ea \right),\ \ \ \ 
K=\left( \ba{cc}
K_{D1}^{} & {K}_{R2}^{} \\
{K}_{R1}^{} & K_{D2}^{}
\ea \right),\ \ \ \ 
J=\left( \ba{cc}
J_{D1}^{} & 0 \\
0 & J_{D2}^{}
\ea \right).
\ee
The expressions of the two-meson loop functions, in terms of the
Mandelstam variable $s$, are 
\be \lb{6e3}
J_{a\bar b,c\bar d}^{}(s,M_{a\bar{b}}^{},M_{c\bar{d}}^{})
=\left\{
\ba{l}
-\frac{i}{2\sqrt{-\lambda_{a\bar b,c\bar d}^{}(s)}}\Big[1-
\frac{1}{\pi}\arctan\Big(\frac{\sqrt{-\lambda_{a\bar b,c\bar d}^{}
(s)}}
{s-(M_{a\bar{b}}^{2}+M_{c\bar{d}}^{2})}\Big)\Big],\\ 
\hspace{5. cm}(M_{a\bar b}^{}-M_{c\bar d}^{})^2<s<
(M_{a\bar b}^{}+M_{c\bar d}^{})^2,\\
\\
+\frac{1}{2\sqrt{\lambda_{a\bar b,c\bar d}^{}(s)}}
\Big[\pm 1+\frac{i}{\pi}\ln\Big
(\frac{\sqrt{s-(M_{a\bar{b}}^{}-M_{c\bar{d}}^{})^2}
+\sqrt{s-(M_{a\bar{b}}^{}+M_{c\bar{d}}^{})^2}}
{\sqrt{s-(M_{a\bar{b}}^{}-M_{c\bar{d}}^{})^2}
-\sqrt{s-(M_{a\bar{b}}^{}+M_{c\bar{d}}^{})^2}}\Big)\Big],\\
\hspace{3. cm} Re(s)>(M_{a\bar{b}}^{}+M_{c\bar{d}}^{})^2,
\ \ \ \ Im(s)=\pm\epsilon,\ \ \epsilon>0,
\ea
\right.
\ee
with
\be \lb{6e4}
\lambda_{a\bar b,c\bar d}^{}(s)=
\Big(s-(M_{a\bar{b}}^{}+M_{c\bar{d}}^{})^2\Big)
\,\Big(s-(M_{a\bar{b}}^{}-M_{c\bar{d}}^{})^2\Big).
\ee
\par
The unitarization operation is realized by the iteration of the
kernel $K$ with the aid of the two-meson loop functions,
\bea \lb{6e5}
\mathcal{T}&=&(1-iKJ)^{-1}K \nonumber \\
&=&\frac{1}{\mathrm{det}(1-iKJ)}
\left( \ba{cc}
(1-iK_{D2}^{}J_{D2}^{})K_{D1}^{}+iK_{R2}^{}J_{D2}^{}K_{R1}^{}
& (1-iK_{D2}^{}J_{D2}^{})K_{R2}^{}+iK_{R2}^{}J_{D2}^{}K_{D2}^{} \\
iK_{R1}^{}J_{D1}^{}K_{D1}^{}+(1-iK_{D1}^{}J_{D1}^{})K_{R1}^{}
& iK_{R1}^{}J_{D1}^{}K_{R2}^{}+(1-iK_{D1}^{}J_{D1}^{})K_{D2}^{}
\ea \right),\nonumber \\
& &
\eea
with
\be \lb{6e6}
\mathrm{det}(1-iKJ)=1-iK_{D1}^{}J_{D1}^{}-iK_{D2}^{}J_{D2}^{}
-K_{D1}^{}J_{D1}^{}K_{D2}^{}J_{D2}^{}
+K_{R1}^{}J_{D1}^{}K_{R2}^{}J_{D2}^{},
\ee
from which one may identify, with (\rf{6e2}), the different
components of $\mathcal{T}$.
\par
We recall that the contact terms $K_R^{}$ and $K_D^{}$ are of
order $1/N_c^{}$ and $1/N_c^2$, respectively. A first
approximation would consist in neglecting $K_D^{}$ in front of
$K_R^{}$. In this case, one has the simplified expressions
\be \lb{6e7}
\mathcal{T}=
\left( \ba{cc}
\mathcal{T}_{D1}^{} & \mathcal{T}_{R2}^{} \\
\mathcal{T}_{R1}^{} & \mathcal{T}_{D2}^{}
\ea \right)\,=\,
\frac{1}{\Big(1+K_{R1}^{}J_{D1}^{}K_{R2}^{}J_{D2}^{}\Big)}
\left( \ba{cc}
iK_{R2}^{}J_{D2}^{}K_{R1}^{} & K_{R2}^{} \\
K_{R1}^{} & iK_{R1}^{}J_{D1}^{}K_{R2}^{}
\ea \right).
\ee
\par
To have a qualitative idea of the behaviors of the scattering
amplitudes, we consider the case of a system made of $c\bar u s\bar d$
constituents ($1\bar 23\bar 4$) in the isospin limit. In the $D1$ channel,
the two mesons are $D^0\bar K^0$ and in the $D2$ channel, they are
$D^+K^-$. Because of isospin symmetry, we have the mass
equalities $M_{1\bar 2}^{}=M_{1\bar 4}^{}$,
$M_{3\bar 4}^{}=M_{3\bar 2}^{}$, and the two-meson thresholds in both
channels are the same: $(M_{1\bar 2}^{}+M_{3\bar 4}^{})^2$.
The value of the contact-term coupling constant $K_{R1}^{}$ has been
estimated from Eq. (\rf{a2e11}), using the expressions of the meson
wave functions, as obtained from the bound state equation  (\rf{2e38}),
normalized according to the condition
$\int_0^\infty dx \varphi_m^{*}(x)\varphi_n^{}(x)=\delta_{mn}^{}$. One
finds $K_{R1}^{}\simeq 7.5(\frac{2\sigma}{\pi})$. The behaviors of the
real and imaginary parts of the scattering amplitudes
$\mathcal{T}_{R1}^{}$ and $\mathcal{T}_{D1}^{}$, as functions of $s$,
are presented in Figs. \rf{f13} and \rf{f14}.
\bfg 
\parbox{8. cm}
{\includegraphics[scale=0.8]{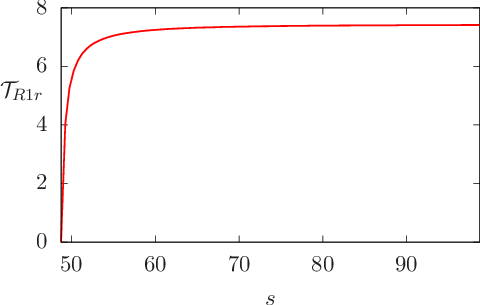}}
\hfill
\parbox{8. cm}
{\includegraphics[scale=0.8]{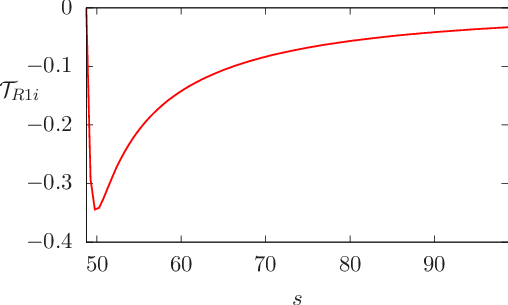}}
\caption{Real and imaginary parts of the recombination scattering
amplitude $\mathcal{T}_{R1}^{}$ in the low energy region (left and
right panels, respectively). The units of $s$ and $\mathcal{T}$ are in
$(\frac{2\sigma}{\pi})\simeq 0.114$ GeV$^2$.} 
\lb{f13} 
\efg
\par
\bfg 
\parbox{8. cm}
{\includegraphics[scale=0.8]{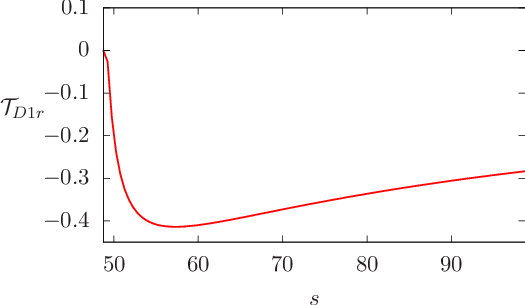}}
\hfill
\parbox{8. cm}
{\includegraphics[scale=0.8]{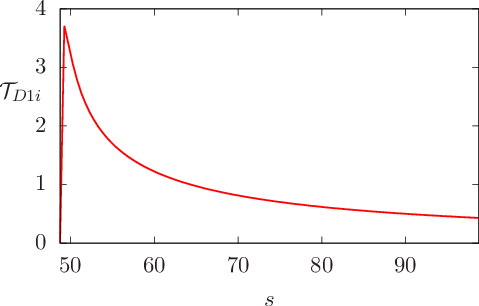}}
\caption{Real and imaginary parts of the direct scattering
amplitude $\mathcal{T}_{D1}^{}$ in the low energy region (left and
right panels, respectively). The units of $s$ and $\mathcal{T}$ are
in $(\frac{2\sigma}{\pi})\simeq 0.114$ GeV$^2$.} 
\lb{f14} 
\efg
\par
It is to be noted that the loop functions $J$ [Eq. (\rf{6e3})]  are
singular at threshold. However, the unitarization operation softens
the behavior of the scattering amplitudes, which then vanish there.
On the other hand, the loop functions vanish at high energies;
therefore, the asymptotic behavior of the scattering amplitudes is
governed by their leading terms in $1/N_c$: The real part of
$\mathcal{T}_{R1}^{}$ tends to $K_{R1}^{}$, while its imaginary
part vanishes; $\mathcal{T}_{D1}^{}$ vanishes with its real and
imaginary parts; had we kept in it the contact term $K_{D1}^{}$,
its real part would asymptotically tend to $K_{D1}^{}$.
\par
The real part of $\mathcal{T}_{D1}^{}$ has a tiny positive value,
with a peak near the threshold, before vanishing again (not visible
with the scale of the figure); this also causes the appearance of
the narrow peak in the imaginary part at the zero of the
real part.
However, these peaks have a purely kinematic origin and do not
correspond to a pole of the scattering amplitude in the complex
plane of $s$ for values of $Re(s)$ above the threshold. More
generally, we have not found resonances in the neighborhood of the 
two-meson thresholds in other cases of quark flavors. This might
be due to the scalar nature of the interaction that is used in the
evaluation of the contact term $K_R^{}$.
\par

\section{Tetraquark bound state equation} \lb{s7}

Existence of tetraquark bound states can be searched for through
the possible occurrence of pole-type singularities in the scattering
amplitude expressions. These correspond to zeros of the denominator
of $\mathcal{T}$ [Eq. (\rf{6e5})], which occur at the zeros of the
determinant (\rf{6e6}). One obtains the equation
\be \lb{7e1}
1-iK_{D1}^{}J_{D1}^{}-iK_{D2}^{}J_{D2}^{}
-K_{D1}^{}J_{D1}^{}K_{D2}^{}J_{D2}^{}
+K_{R1}^{}J_{D1}^{}K_{R2}^{}J_{D2}^{}=0.
\ee
\par
To have a qualitative idea of the content of this equation, we
consider the case where the meson masses are the same in both
channels $D1$ and $D2$: $M_{1\bar 2}^{}=M_{1\bar 4}^{}$,
$M_{3\bar 4}^{}=M_{3\bar 2}^{}$. In that case, the two meson loop
functions are equal, $J_{D1}^{}=J_{D2}^{}\equiv J_D^{}$ and similarly
for the direct-channel contact terms,
$K_{D1}^{}=K_{D2}^{}\equiv K_D^{}$; generally, by unitarity, one has
$K_{R1}^{}=K_{R2}^{}\equiv K_R^{}$. Equation (\rf{7e1}) becomes
\be \lb{7e2}
\Big(1-i(K_R^{}+K_D^{})J_D^{}\Big)\,
\Big(1+i(K_R^{}-K_D^{})J_D^{}\Big)=0.
\ee
$K_D^{}$ is of order $1/N_c^2$ and is expected to be smaller in
modulus than $|K_R^{}|$, which is of order $1/N_c^{}$. Within this
situation, and noticing that $J_D^{}$ below threshold is imaginary
[Eq. (\rf{6e3})], one finds that whatever the sign of $K_R^{}$ is,
Eq. (\rf{7e2}) has generally a bound state solution in $s$.
This conclusion is not modified in the more general case of
unequal meson masses in the two channels $D1$ and $D2$.
\par
Neglecting $K_{D1}^{}$ and $K_{D2}^{}$ in Eq. (\rf{7e1}) and after
introducing the expressions of $J_{D1}^{}$ and $J_{D2}^{}$
[Eq. (\rf{6e3})], the bound state equation takes the form 
\bea \lb{7e3}
\sqrt{16(-\lambda_{D1}^{})(-\lambda_{D2}^{})}&-&
K_R^2\Big[1-\frac{1}{\pi}\arctan\Big(\frac{\sqrt{-\lambda_{D1}^{}}}
{(s-(M_{1\bar 2}^2+M_{3\bar 4}^2))}\Big)\Big]\nonumber \\  
& &\times 
\Big[1-\frac{1}{\pi}\arctan\Big(\frac{\sqrt{-\lambda_{D2}^{}}}
{(s-(M_{1\bar 4}^2+M_{3\bar 2}^2))}\Big)\Big]=0,
\eea
where $\lambda_{D1}^{}=\lambda_{1\bar 2,3\bar 4}^{}$ and
$\lambda_{D2}^{}=\lambda_{1\bar 4,3\bar 2}^{}$ [Eq. (\rf{6e4})].
\par
The value of $K_R^{}$, which is determined from Eq. (\rf{a2e11}),
turns out to be sensitive to the flavor content of the system
under study. We defer to a separate work the detailed analysis
of this equation.
\par
One main conclusion can be drawn, however, at the present stage.
The coupling constant $K_R^{}$ in Eq. (\rf{7e3}), being of
order $1/N_c^{}$, vanishes in the extreme limit
$N_c\rightarrow\infty$ and hence the existing tetraquark
approaches the nearest two-meson threshold and disappears there.
This confirms Weinberg's conjecture \cite{Weinberg:2013cfa} that
if multiquark states exist, then they can only appear in
nonleading terms in $1/N_c^{}$.
This is in contrast with the ordinary meson spectra, which remain
almost unchanged in the large-$N_c^{}$ limit.
\par

\section{Concluding remarks} \lb{s8}

The main result of the present work is the demonstration, in
two-dimensional QCD, on the basis of an infrared regularization
scheme and in the case of four different quark flavors, of the
finiteness of the meson scattering amplitudes up to order $1/N_c^2$.
The possibility of their unitarization allows us to investigate
tetraquark spectroscopic problems and their internal dynamics in
a more explicit framework. The results are free of external empirical
parameters. The parameters of the theory are the coupling constant,
or equivalently the string tension, which fixes the mass scale of
the theory, and the quark masses. The equations that are obtained
are fully relativistic and allow probing various energy regions
with appropriate expansions.
\par
One of the striking features of the infrared limiting procedure is
the possibility of transforming a four-quark scattering process into
an effective two-meson scattering process with contact-type
interactions, representing here short-range interactions, generating
unitarity diagrams and where the physical outcomes are expressible
in terms of the meson wave functions and propagators.
This is what one expects from QCD, however its two-dimensional
version displays here explicitly the corresponding dynamical
mechanism. The latter is dragged by the lowest-order quark-exchange
diagrams in $1/N_c^{}$, which, through unitarization, provide the
dominant contributions.
\par
The bound state equation (\rf{7e3}) obtained from the on-mass shell
scattering amplitudes does not, however, provide us with complete
information about the bound state wave functions. Apart from the
mass of the bound state, it gives information, by means of the
residues of the bound state pole in the four channels, about the
couplings of the
bound state to the two-meson clusters, which are represented by
particular moments of the wave functions. To obtain full information
about the wave functions, one has to go back to the integral
equations (\rf{4e14}) and (\rf{4e15}) satisfied by the Green's
functions and convert them into bound state equations through their
homogeneous parts. Here, again, one faces the infrared divergence
problem, which cannot, however, be solved in a parallel way with
the procedure adopted for the scattering amplitudes, because of the
differences in their structures.
The way to reach infrared finite expressions is to pass to the
instantaneous reduction of the equations and of the wave functions.
This was the case for the quark-antiquark systems: it is the
instantaneous wave function that is finite and leads to observable
quantities. The problem, however, is more complicated in the present
many-body case; simple global factorization of the wave function
is no longer sufficient to reach that aim, because that initial
ansatz is not self-reproducible. The correct method necessitates
more complicated transformations, which gradually bring the
above equations into instantaneously reduced finite forms. This
aspect of the problem will be presented in a separate work.
\par
The present study focused on the case of four different quark
flavors, which offers the simplest theoretical framework, avoiding
mixing problems with the ordinary meson sectors. The case
containing hidden flavors, like $c\bar c$ or $b\bar b$,
involves additional contributions, represented by quark-exchange
diagrams in the direct scattering channels (cf. Refs.
\cite{Callan:1975ps,Einhorn:1976ax,Einhorn:1977bg}), leading to
mixings with ordinary meson sectors. To establish integral equations,
one needs the inclusion of additional kernels or couplings with
two-quark sectors, which leads to the enlargement of the space of
the initially introduced Green's functions. 
This aspect of the problem requires a more specific treatment.
Nevertheless, one can consider direct calculations of the scattering
amplitudes with such diagrams; their main contribution starts at
order $1/N_c^{}$ and, their finite part, in the limit of vanishing
cutoff, reduces to contact terms plus meson-exchange contributions.
Approximating also the
latter by contact terms, the whole contribution can be incorporated
in the kernels $K_D^{}$, previously introduced [Eq. (\rf{6e2})], and
through them into the unitarized scattering amplitudes. The only
novelty is that the kernels $K_D^{}$ are now of order $1/N_c^{}$
and enter into competition with $K_R^{}$. In particular, contrary
to the case of $K_R^{}$, the sign of $K_D^{}$, in case it is
dominant, plays a crucial role for the existence of bound states.
\par
The particular case when two quarks or two antiquarks have identical
flavors, like $cc$ or $bb$, also merits attention. Here,
recombination and direct channels are identical and, therefore, one
has a single-channel problem. 
\par
The ability of the two-dimensional theory
to produce confinement and a linear meson spectrum at high energies,
with a linear potential in coordinate space in the nonrelativistic
limit (the latter demonstrated more easily in the axial gauge),
is an indication that some of the predictions obtained in
multiquark sectors might also be relevant in four dimensions.
Two-dimensional models cannot, however, describe rotational motion,
or transverse motion in light-cone bases. One is limited with
sectors of pseudoscalar or scalar mesons with their radial
excitations. Nevertheless, the mechanism of confinement and the
explicit disappearance of infrared divergences in physical
quantities in two dimensions might provide us with hints about
similar mechanisms in four dimensions. 
Another point of relevance is the possibility of extracting from
the spectrum of tetraquarks in two dimensions a more detailed
description of their structure and their properties.
\par


\begin{acknowledgments}
The author thanks Wolfgang Lucha, Dmitri Melikhov and Bachir
Moussallam for stimulating discussions on the subject.
This research has received financial support from the EU research
and innovation programme Horizon 2020, under Grant Agree\-ment
No. 824093.
\end{acknowledgments}


\appendix

\section{} \lb{sa1}

In this Appendix, we evaluate the contributions of the gluon
propagators to the various types of Feynman diagrams.
\par
We first consider the generic case of a gluon
propagator acting on a wave function $f$ by means of a convolution
operation. We adopt for the gluon propagator the regularization
method of introducing a small mass term $\lambda$ in the denominator       
[Eq. (\rf{2e14})]. The structure of the resulting integral is of
the type
\be \lb{a1e1}
I=\int_c^r\frac{dk}{2\pi}\frac{f(k+p)}{k^2+\lambda^2},
\ee
where $c$ and $r$ are bounds coming from the spectral conditions
satisfied by the wave function [cf. example of Eq. (\rf{2e33})].
The leading effect of the singularity of the gluon propagator is
obtained by isolating the function $f(k+p)$ with its value at the
position of the singularity, i.e., $f(p)$, and factorizing it outside
the integral, obtaining
\be \lb{a1e2}
I=f(p)\int_c^r\frac{dk}{2\pi}\frac{1}{k^2+\lambda^2}+
\int_c^r\frac{dk}{2\pi}\frac{\big(f(k+p)-f(p)\big)}{k^2}.
\ee
In the second integral, the mass term has been suppressed, since
the integral is finite (with the principal value prescription).
The first integral can be evaluated,
\be \lb{a1e3}
\int_c^r\frac{dk}{2\pi}\frac{1}{k^2+\lambda^2}=
\frac{1}{2\pi\lambda}\Big(\arctan(\frac{r}{\lambda})
-\arctan(\frac{c}{\lambda})\Big).
\ee
In the limit $\lambda\rightarrow 0$, one has
\be \lb{a1e4}
{\int_c^r\frac{dk}{2\pi}\frac{1}{k^2+\lambda^2}}_{\stackrel{{
\displaystyle\ =\ }}{\lambda\rightarrow 0}}\frac{1}{4\lambda}
(\varepsilon(r)-\varepsilon(c))-\frac{1}{2\pi}(\frac{1}{r}-
\frac{1}{c}),
\ee
where $\varepsilon(x)=$sgn$(x)$.
\par
It is seen that the singularity shows up only when $r$ and $c$
are of opposite signs. In case one is interested in the leading
term, it is then sufficient to replace the bounds by $\pm\infty$,
according to the signs of $r$ and $c$.
\par
In many Feynman diagrams, some of the internal quark lines may
not be subjected to the spectral conditions and may involve
wider domains of integration, with negative $p_-^{}$ components.
The latter, following the structure of the quark propagators
[Eq. (\rf{2e18})], change the sign of the $i\epsilon$ factor of
the denominator and modify the evaluation of the corresponding
Cauchy integral related to the $p_+^{}$ component. It turns out that
the resulting bounds of integration in integrals of the type of
(\rf{a1e1}) become now of the same sign and therefore do not
contribute to the leading singular terms. For the evaluation
of the latter, one may stick from the start to bounds satisfying
everywhere the spectral conditions.
\par
We next consider the case of several gluon propagators
that appear in multiple integrals, in the direct channels, in
convolution with each other.
We concentrate on the treatment of the singular part of  
such integrals, since it is only that part that contributes in
the finite infrared limit of multigluon diagrams that we meet
in the case of direct processes of Sec. \rf{s5}.
\par
A generic form of the integrals with $n$ ($n\ge 2$) gluon
propagators, after changes of variables by translation, is the
following:
\be \lb{a1e5}
I_n^{}(q)=\int \Big(\prod_{i=1}^{n-1}\frac{dk_i^{}}{2\pi}
\frac{1}{(k_i^2+\lambda^2)}\Big)\frac{1}{\big((q-k_1^{}-k_2^{}
-\cdots -k_{n-1}^{})^2+\lambda^2\big)},
\ee
where $q$ is an external momentum, typically a momentum transfer
between an ingoing meson and an outgoing one, of the form
$q=r_{1\bar 2}'-r_{1\bar 2}^{}$, for instance. The variables
$k_i^{}$ ($i=1,2,\ldots,n-1$) are themselves differences of
momenta of neighboring quark or antiquark lines and, therefore,
in reference to the integral (\rf{a1e4}), their bounds, according to the
spectral conditions of the wave functions and their related
propagators, contain the point 0, with $r_i^{}>0$ and $c_i^{}<0$.
It is only in the last propagator of Eq. (\rf{a1e5}) that the bounds may
play a crucial role.
\par
Without taking into account the question of the bounds and making
the calculation of the integral (\rf{a1e5}) by recursion, one finds
$I_n^{}(q)=\frac{1}{(2\lambda)^{n-1}}\frac{n}{(q^2+(n\lambda)^2)}$.
One notices the modification of the infrared parameter with the
accompanying number of gluon propagators. In general, the diverging
factor $1/(2\lambda)^{n-1}$ is canceled by a similar damping
factor coming from the quark propagators and their integration.
\par
Considering now the bounds of the variables $k_i^{}$
and taking for $q^2$ large values,
one may reach domains of momenta where the sum $\sum_ik_i^{}$
cannot cancel the momentum transfer $q$ and, therefore,
in these circumstances, the last
propagator does not contribute to the leading singularity of the
integral. Since the other singularities of $k_i^{}$ are located
at the value $0$, the final contribution of the last
propagator would be $1/(q^2+\lambda^2)$, while the other
integrals would give the factor $1/(2\lambda)^{n-1}$.
This result contradicts the one obtained above without taking
into account the integration bounds and for which the behavior of
the integral for large values of $q^2$, without the diverging factor,
would be $n/q^2$ and not $1/q^2$. It is obvious that the latter
result is the correct one, since it takes into account the existence
of the integration bounds. How does one reconcile, however, the two
results for small values of $q^2$, since in these domains the former
result seems also correct, the bounds no longer playing a decisive
role? The answer is related to the property that for small values of
$q^2$ and in the limit of vanishing $\lambda$, one has the
equivalence relation
\be \lb{a1e6}
\Big(\frac{n\lambda}{q^2+(n\lambda)^2}\Big)
_{\stackrel{{\displaystyle\ =\ }}
{\lambda\rightarrow 0}}\,
\frac{1}{\pi}\delta(q)_{\stackrel{{\displaystyle\ =\ }}
{\lambda\rightarrow 0}}\Big(\frac{\lambda}{q^2+\lambda^2}\Big).
\ee
We therefore conclude that the expression of the integral (\rf{a1e5})
is
\be \lb{a1e7}
I_n^{}(q)=\frac{1}{(2\lambda)^{n-1}}\frac{1}{(q^2+\lambda^2)}.
\ee
\par
We finally add the following remark. Apart from the diverging or
singular expressions that we have found in the above calculations,
one also finds finite nonsingular expressions, which do not disappear
in the infrared limit $\lambda\rightarrow 0$ and which arise from
the presence of various numerators in the integrals (\rf{a1e5}),
depending on the total momentum transfer, which we have factorized
for simplicity. The existence of such terms should be taken into
account, at least qualitatively, for the physical interpretation
of the results. Generally, they produce, in the mesonic sectors,
contact-type interactions.
\par

\section{} \lb{sa2} 

This Appendix is devoted to the presentation of the details of the
calculations of the various contributions to the meson-meson
scattering amplitudes to order $1/N_c^2$ in the limit
$\Lambda\rightarrow\infty$ (vanishing of the infrared cutoff).
\par
We first consider the recombination channel scattering amplitude,
represented by the diagrams of Fig. \rf{f6}. To simplify the formulas,
we shall adopt the following notations,
\be \lb{a2e1}
a_i^{\prime}(k_{i-}^{})=\frac{m_i^{\prime 2}}{2k_{i-}^{}},\ \ \ \ \
a_i^{}(k_{i-}^{})=\frac{m_i^{2}}{2k_{i-}^{}},\ \ \ \ \ 
i=1,\bar 2,3,\bar 4,
\ee
$a_i^{\prime}$ being quantities that appear in the quark propagators
[Eq. (\rf{2e18})] and the relation between
$m^{\prime 2}$ and $m^2$ is given in Eq. (\rf{2e19}).
We shall often omit, when no ambiguities arise, the mentioning
of their momentum, the latter being easily recognized from the
corresponding graphs. The convention of Eq. (\rf{4e1})
($S_{a,-}^{}\rightarrow S_a^{}$) will be adopted. Furthermore,
according to the analysis of the spectral conditions (cf. Sec.
\rf{s5}), coming from the wave functions, and their incidence on the
internal quark lines in the limit $\Lambda\rightarrow\infty$, the
quarks and antiquarks of the loops are assumed to satisfy
individually those conditions. This has a simplifying consequence
on the expressions of the quark propagators, concerning the signs
of the $-$ components of the momenta.
\par
Of particular help is the simplest formula for two-propagator
integrations:
\bea \lb{a2e2}
& &\int\frac{dk_+^{}}{2\pi}S_1^{}(k)S_2(k-r_{1\bar 2}^{})
=\int\frac{dk_+^{}}{2\pi}
\frac{ik_-^{}}
{(k^2-|k_-^{}|\Lambda-m_1^{\prime 2}+i\epsilon)}
\frac{i(k_-^{}-r_{1\bar 2-}^{})}
{((k-r_{1\bar 2}^{})^2-|k_-^{}-r_{1\bar 2-}^{}|\Lambda
-m_2^{\prime 2}+i\epsilon)}
\nonumber \\
& &\ \ \ =-\frac{i}{4}\frac{\theta(k_-^{}(r_{1\bar 2-}^{}-k_-^{}))}
{\Big(r_{1\bar 2+}^{}-\Lambda
-a_1^{\prime}(k_-^{})-a_2^{\prime}(r_{1\bar 2-}^{}-k_-^{})\Big)}.
\eea
\par
Diagram (a) of Fig. \rf{f6} contains the following integral:
\bea \lb{a2e3}
\widetilde I_{(a)}^{}&=&\int\frac{dk_+^{}}{2\pi}S_1^{}(k)
S_2(k-r_{1\bar 2}')
S_3^{}(r_{3\bar 2}^{}-r_{1\bar 2}'+k)S_4^{}(k-r_{1\bar 4}^{})
\nonumber \\
&=&\frac{i^4}{16}\int\frac{dk_+^{}}{2\pi}
\frac{1}{(k_+^{}-a_1'-\frac{\Lambda}{2}+i\epsilon)}
\frac{1}{(k_+^{}-r_{1\bar 2+}'+a_2'+
\frac{\Lambda}{2}-i\epsilon)}
\nonumber \\
& &\ \ \ \ \ \ \times 
\frac{1}{(r_{3\bar 2+}^{}-r_{1\bar 2+}'+k_+^{}-a_3'
-\frac{\Lambda}{2}+i\epsilon)}
\frac{1}{(k_+^{}-r_{1\bar 4+}^{}+a_4'
+\frac{\Lambda}{2}-i\epsilon)}.
\eea
(The full diagram contribution, including the external wave functions
and the integration over $k_-$, will be designated by $I_{(a)}$.)
After integrating with respect to $k_+^{}$ with the Cauchy theorem
and expanding terms with respect to $1/\Lambda$, one finds
\be \lb{a2e4}
\widetilde I_{(a)}^{}=\frac{2i}{16\Lambda^3}
\frac{[1-(r_+-a_1'-a_{\bar 2}'-a_3'-a_{\bar 4}')/(2\Lambda)]}
{\scriptstyle{(1-(r_{1\bar 2+}'-a_1'-a_{\bar 2}')/\Lambda)
(1-(r_{3\bar 4+}'-a_3'-a_{\bar 4}')/\Lambda)
(1-(r_{1\bar 4+}{}-a_1'-a_{\bar 4}')/\Lambda)
(1-(r_{3\bar 2+}^{}-a_3'-a_{\bar 2}')/\Lambda)}}.         
\ee
Notice that we have not expanded the denominator in terms of
$1/\Lambda$. The reason is that each factor of the denominator
is equal to the contribution of the finite corrective part of
the corresponding wave function, according to Eq. (\rf{2e47}),
the latter contributing at the end by a multiplicative factor.
This shows that the finite corrective part of the wave functions
will be canceled and will not contribute to the result. 
\par
Equation (\rf{a2e4}) shows that $\widetilde I_{(a)}^{}$, after
multiplication by the factor $\Lambda^4$ coming from the four
multiplicative external wave functions, leads to a linearly divergent
quantity in $\Lambda$, plus a finite part represented by the
contribution of the second term of the numerator of the fraction.
\par
The integrals coming from diagrams (b) and (c) of Fig. \rf{f6}
can be calculated in a similar way. Here, one has two loop momenta,
$k$ and $k'$, and six quark propagators. The integration with
respect to the $+$ component of each of them will concern three
quark propagators. One then considers the contribution of the
gluon propagator. We shall concentrate for the moment on the
singular part of it, whose contribution is taken into account
with the prescriptions of Eqs. (\rf{a1e1})-(\rf{a1e4}), or
(\rf{2e34}) and (\rf{2e35}). The singular part simply equates the
components $k_-^{}$ and $k_-'$. We shall designate the corresponding
integrals by $\widetilde I_{(b/c)}^{}$. The contributions coming
from the regular part of the gluon propagator, having a different
structure, will be treated separately. One finds
\be \lb{a2e5}
\widetilde I_{(b)}^{}=\widetilde I_{(c)}^{}=
-\frac{i}{16\Lambda^3}
\frac{1}
{\scriptstyle{(1-(r_{1\bar 2+}'-a_1'-a_{\bar 2}')/\Lambda)
(1-(r_{3\bar 4+}'-a_3'-a_{\bar 4}')/\Lambda)
(1-(r_{1\bar 4+}{}-a_1'-a_{\bar 4}')/\Lambda)
(1-(r_{3\bar 2+}^{}-a_3'-a_{\bar 2}')/\Lambda)}}.         
\ee
\par
One observes that the diverging part of $\widetilde I_{(a)}^{}$ is
canceled by the sum of of $\widetilde I_{(b)}^{}$ and
$\widetilde I_{(c)}^{}$:
\be \lb{a2e6}
\widetilde I_{(a)}+\widetilde I_{(b)}^{}+\widetilde I_{(c)}^{}=
-\frac{i}{16\Lambda^4}
\big(r_+^{}-a_1'-a_{\bar 2}'-a_3'-a_{\bar 4}'\big).
\ee
(There is no need now to keep the denominators of Eqs. (\rf{a2e4})
and (\rf{a2e5}), since it is only the diverging parts of the wave
functions that can contribute to maintain the quantity different
from zero.)
\par
One also has to include, according to Eq. (\rf{a1e4}), with the
diverging contribution of the gluon propagator, a finite part
coming from the integration bounds. The latter contribution
transforms, as in the case of the wave equation (\rf{2e38}),
$m^{\prime 2}$ into $m^2$ [Eq. (\rf{2e19})]. One obtains
\be \lb{a2e7}
\widetilde I_{(a)}+\widetilde I_{(b)}^{}+\widetilde I_{(c)}^{}=
-\frac{i}{16\Lambda^4}
\big(r_+^{}-a_1^{}-a_{\bar 2}^{}-a_3^{}-a_{\bar 4}^{}\big).
\ee
The term in the parentheses in Eq. (\rf{a2e7})
can be divided into four parts,
\be \lb{a2e8}
r_+^{}-a_1^{}-a_{\bar 2}^{}-a_3^{}-a_{\bar 4}^{}=\frac{1}{2}\Big[\,
(r_{1\bar 2 +}'-a_1^{}-a_{\bar 2}^{})
+(r_{3\bar 4 +}'-a_3^{}-a_{\bar 4}^{})
+(r_{1\bar 4 +}-a_1^{}-a_{\bar 4}^{})
+(r_{3\bar 2 +}-a_3^{}-a_{\bar 2}^{})\,\Big],
\ee
each representing a part of the wave equation
satisfied by the corresponding wave function having the same quark
content. Equation (\rf{a2e7}) becomes
\be \lb{a2e9}
\widetilde I_{(a)}+\widetilde I_{(b)}^{}+\widetilde I_{(c)}^{}=
-\frac{i}{32\Lambda^4}
\Big[\,(r_{1\bar 2 +}'-a_1^{}-a_{\bar 2}^{})
+(r_{3\bar 4 +}'-a_3^{}-a_{\bar 4}^{})
+(r_{1\bar 4 +}-a_1^{}-a_{\bar 4}^{})
+(r_{3\bar 2 +}-a_3^{}-a_{\bar 2}^{})\,\Big].
\ee
\par
We next consider the contribution of the regular part of the gluon
propagator. Considering, for definiteness, diagram (b) of Fig.
\rf{f6}, we notice that we have now integrals with respect to
$k_-^{}$ and $k_-'$, with the gluon propagator momentum $q_-$ equal
to $(k_-'-k_-^{})$. Considering $k_-^{}$ as fixed, the integration
with respect to $k_-'$ will concern the wave functions
$\varphi_{1\bar 2}^{}(r_{1\bar 2-}',k_-')$ and 
$\varphi_{3\bar 2}^{}
(r_{3\bar 2-}^{},r_{3\bar 4-}'-r_{1\bar 4-}^{}+k_-')$,
which we shall shorten by the notations $\varphi_{1\bar 2}^{}(k_-')$
and $\varphi_{3\bar 2}^{}(k_-')$. A similar shorthand notation is
also used for the two other wave functions,
$\varphi_{1\bar 4}^{}(k_-^{})$ and $\varphi_{3\bar 4}^{}(k_-^{})$. 
\par
After having integrated with respect to $k_+^{}$ and $k_+'$, one
remains with four quark propagators, each having the damping
factor $(-\Lambda^{-1})$. The global damping factor $\Lambda^{-4}$ is
then canceled by the diverging factor $\Lambda^4$, coming from
the external wave functions. The contribution of the regular part
of the gluon propagator is then
\bea \lb{a2e10}
I_{(b),gl}^{}&=&16\pi^2 (-2\sigma)\int\frac{dk_-^{}}{2\pi}
\varphi_{1\bar 4}^{}(k_-^{})\varphi_{3\bar 4}^{}(k_-^{})
\int\frac{dk_-'}{2\pi}\frac{i}{(k_-'-k_-^{})^2}
\Big(\varphi_{1\bar 2}^{}(k_-')\varphi_{3\bar 2}^{}(k_-')
-\varphi_{1\bar 2}^{}(k_-^{})\varphi_{3\bar 2}^{}(k_-^{})\Big).
\nonumber \\
& &
\eea
The contribution of diagram (c) has a similar expression, obtained
by appropriate interchange of indices.
\par
The integration bounds of the variable $k_-^{}$ in (\rf{a2e9})
and of $k_-^{}$ and $k_-'$ in (\rf{a2e10}) have to respect the
spectral conditions of the wave functions. However,
each wave function depends, in general, on a different total
momentum $r_{a\bar b-}^{}$ and on a different quark momentum,
$k_{a-}^{}$, with a combined dependence upon the variable
$x_{a\bar b}^{}=k_{a-}^{}/r_{a\bar b-}^{}$. Converting for instance
the integration momentum variable $k_-^{}$ of Fig. \rf{f6}(a)
into the variable $x_{1\bar 2}^{}$, one is left in the other wave
functions with the presence of the ratios
$r_{1\bar 2-}'/r_{1\bar 4-}$,
$r_{1\bar 2-}'/r_{3\bar 2-}$ and $r_{1\bar 2-}'/r_{3\bar 4-}'$,
which, generally are not equal to 1. The domain of integration of
$x_{1\bar 2}^{}$ is $[0,1]$, but taking into account the values
of the latter ratios, which themselves may change according to the
external conditions, it is restricted to a smaller domain.
\par
The amplitudes resulting from expressions (\rf{a2e9}) and (\rf{a2e10})
are equivalent to generalized momen\-tum-dependent contact
interactions, for they do not induce long-range type interactions,
remaining finite in various momentum limits. However, 
the exact analytic calculation of the restricted domain of integration
and of its momentum dependence is a rather complicated task, which,
in its general case demands detailed numerical evaluations.
Furthermore, we will be interested in the present work by the unitary
summation of the low-order amplitudes to take into account the
most important higher-order effects. This involves integrations of
meson-loop momenta; the presence of momentum transfer variables
inside the wave functions, which generally are known numerically,
does not allow the realization of such a summation in analytic form.  
It is therefore preferable to stick, at the present stage, to the
scalar approximatation of the contact interactions.
This can be done by adopting for the external momentum ratios the
following simplifications: $r_{1\bar 2-}'/r_{1\bar 4-}=1$,
$r_{1\bar 2-}'/r_{3\bar 2-}=1$ and $r_{1\bar 2-}'/r_{3\bar 4-}'=1$.
In terms of the Mandelstam variables, they would correspond  to
$t=0$ and $u=0$ and in the equal-mass case of mesons to $s=4M^2$,
where $M$ is the mass of the external mesons. For the case of unequal
masses, the above choices of momentum ratios may not correspond to
physical values. However, that simplification concerns only the
internal domain of integration of the wave functions and not the
external momenta, which might still appear in other multiplicative
expressions.
\par
Another remark concerns the case of bound states. For values of the
variable $s$ lying below two-meson thresholds, some of the
momentum ratios mentioned above may become complex. This is only
an apparent difficulty, since the analyticity property in $s$, which
should allow continuation to the bound state domain, concerns the
scattering amplitude itself and not intermediate or auxiliary
variables. One therefore has first to calculate the scattering
amplitude in its physical domain and then proceed to its continuation
to the bound state region.
\par
Taking into account the abovementioned simplifications, 
the final expression for the contribution of the diagrams of
Fig. \rf{f6} is,
\bea \lb{a2e11}
I&=&I_{(a)}+I_{(b)}+I_{(c)}\equiv iN_c^{}K_{R1}^{}\nonumber \\
&=&i\,2\pi 
\Big\{\,\int_0^1 dx \varphi_{1\bar 2}^{}(x)\varphi_{3\bar 4}^{}(x)
\varphi_{1\bar 4}^{}(x)\varphi_{3\bar 2}^{}(x)\nonumber \\
& &\ \ \ \ \ \ \ \times \Big[\,(M_{1\bar 2}^2+M_{3\bar 4}^2
+M_{1\bar 4}^2+M_{3\bar 2}^2)
-2\frac{(m_1^{2}+m_3^{2})}{x}
-2\frac{(m_2^{2}+m_4^{2})}{(1-x)}\,\Big]\,\Big\}
\nonumber \\  
& &-i\,4\pi (\frac{2\sigma}{\pi})
\Big\{\,\int_0^1 dx \varphi_{1\bar 4}^{}(x)\varphi_{3\bar 4}^{}(x)
\int_0^1 dx'\frac{1}{(x'-x)^2}
\Big(\varphi_{1\bar 2}^{}(x')\varphi_{3\bar 2}^{}(x')
-\varphi_{1\bar 2}^{}(x)\varphi_{3\bar 2}^{}(x)\Big)\nonumber \\
& &\ \ \ \ \ \ \
+\int_0^1 dx \varphi_{1\bar 2}^{}(x)\varphi_{3\bar 2}^{}(x)
\int_0^1 dx'\frac{1}{(x'-x)^2}
\Big(\varphi_{1\bar 4}^{}(x')\varphi_{3\bar 4}^{}(x')
-\varphi_{1\bar 4}^{}(x)\varphi_{3\bar 4}^{}(x)\Big)\,\Big\},
\eea
where $M_{a\bar b}^2$ are the external meson masses squared.
A similar expression is also valid for the
recombination amplitude $R_2^{}$, which is obtained from
$R_1^{}$ by the interchanges
$(1\bar 2)(3\bar 4)\leftrightarrow(1\bar 4)(3\bar 2)$.
\par
We next consider the direct channel scattering amplitude and
the corresponding parts, represented in Eqs. (\rf{5e2})-(\rf{5e7}).
\par
The way of calculating expression (c), represented for one of its
parts in Fig. \rf{f8}, has been explained in the main text. The last
technical point that remains is the evaluation of the integral with
respect to the loop momentum $l_+^{}$. The momentum $l$ appears in
the meson propagators and in two quark propagators in each
recombination diagram, after the integrations with respect to the
other momenta have been done. We recall that the limit
$\Lambda\rightarrow\infty$ should not be taken before the last
integration is done (at least for the quantities depending on $l$).
The integration with respect to $l_+^{}$ receives two types of
contribution: a first one comes from the meson propagators, a second
one from the quark propagators.
\par
The meson propagators do not contain $\Lambda$. In analogy with
(\rf{a2e1}), we introduce, for the meson having the flavor and
excitation quantum numbers $(a\bar b)$ and $n$, the following
notation:
\be \lb{a2e12}
A_{a\bar b,n}^{}(l_{a\bar b-}^{})=\frac{r_{a\bar b,n}^2}
{2l_{a\bar b-}^{}},
\ee
where $l_{a\bar b}^{}$ is a generic momentum, carried by the meson,
and $r_{a\bar b,n}^2$ is the mass squared of the meson.
The contribution coming from one of the meson propagators produces
with the other meson propagator the term 
\be \lb{a2e13}
J_{1\bar 4,3\bar 2}^{}(r,l_-^{})=\frac{i}{4l_-^{}(r_-^{}-l_-^{})
\big[r_+^{}-A_{1\bar 4,p}^{}(l_-^{})
-A_{3\bar 2,q}^{}(r_-^{}-l_-^{})+i\epsilon\big]},  
\ee
which is the two-meson once-integrated loop contribution.
The $\Lambda$ dependence of the quark propagators of the
recom\-bi\-nation-diagram parts is not modified by this
contribution. Taking then the limit $\Lambda\rightarrow\infty$,
one finds a finite contribution for each recombination part
on both sides of the meson loop contribution, given by
expressions of the type of Eq. (\rf{a2e11}), where the external
momenta contained in the meson wave functions have to be fixed
according to the figure. In particular, the momenta of the wave
functions coming from the internal meson propagators depend on
$l_-^{}$. Because of the spectral conditions that are satisfied
at the quark-propagator level, the integration bounds of $l_-^{}$
are $0$ and $r_-^{}$. $r_+^{}$ is fixed by the external ingoing or
outgoing meson momenta; for instance,
$r_+^{}=r_{1\bar 2+}'^{}+r_{3\bar 4+}'^{}=r_{1\bar 2,n'}^2/
(2r_{1\bar 2-}')+r_{3\bar 4,m'}^2/(2r_{3\bar 4-}')$.
$l_-^{}$ itself enters in the integration bounds of the internal
momenta of the recombination-channel parts of the effective diagram.
In general, one has an off-energy type integration.
\par
The contribution coming from the quark propagators during the
integration of $l_+^{}$ has two effects: (i) the degree of the damping
factor in $\Lambda$ decreases simultaneously by one power in each
recombination part on both sides of the meson propagators; (ii) each
meson propagator catches one damping power in $\Lambda$. The
global damping power in $\Lambda$ remains therefore the same
($\Lambda^{-8}$). When the limit $\Lambda\rightarrow\infty$ is
taken, the quark and meson propagators disappear in favor of the
factor $\Lambda^{-8}$.
One remains, at leading order, with the difference of two equal
quantities, which reduces to zero.
Therefore, expression (c) produces in the limit
$\Lambda\rightarrow\infty$ only the effective two-meson loop
contribution between the two effective recombination contact
terms (Fig. \rf{f9}).
\par
Expressions (a), (b), (d) and (e) of Eqs. (\rf{5e2}), (\rf{5e3}),
(\rf{5e4}), (\rf{5e6}), and (\rf{5e7}) can be evaluated with the
same methods and power countings as for the cases met before.
The calculation of integrals involving several
gluon propagators is done in Appendix \rf{sa1} [Eqs. (\rf{a1e5}) and
(\rf{a1e7})]. The only novelty with respect to the previous cases is
that in the corresponding diagrams one finds at the end
a finite quantity, factorized by a gluon propagator carrying the
momentum transfer of the process, which, therefore, is no longer
involved in the internal integrations [cf. Eq. (\rf{a1e7})].
Such a term, considered separately, would produce
long-range forces in coordinate space. It is therefore necessary
to check, in each category of diagrams mentioned above, the
cancelation of such terms or their reduction to short-range
type expressions. 
\par
The detailed calculation of the contributions of the corresponding
diagrams shows that the sum of contributions of terms factorized
with an external-type gluon propagator in each category,
(a), (b), (d), (e), is equal to zero. As emphasized at the end of
Appendix \rf{sa1}, one still has contributions coming from finite
nonsingular terms, which play the role of contact-type
interactions in the effective meson theory. Because of the big
number of such contributions, the precise evaluation of the
expression of the contact term in terms of meson wave functions
requires lengthy calculations that go beyond the limitations
of the present work. We shall be content with admitting, in addition
to the unitarity diagrams resulting from the category (c) above,
the existence of such an interaction in the direct channels.
\par
Finally, let us mention that, according to the way of calculation,
one may find in some particular diagrams, such as in category (d),
instead of a gluon propagator, a delta function of the momentum
transfer. In such cases, one may use the equivalence relation
(\rf{a1e6}), to convert the delta function into a propagator in
the infrared limit $\lambda\rightarrow 0$.
\par


\end{document}